\documentclass[3p,authoryear,a4paper]{elsarticle}
\usepackage{graphicx}
\usepackage{slashbox}
\usepackage{color}
\usepackage{dsfont}
\usepackage{amsmath}
\usepackage{amssymb}
\usepackage{amsfonts}
\usepackage{amsbsy}
\usepackage{subcaption}
\usepackage{amsthm}
\usepackage{array}
\usepackage{booktabs} 
\usepackage{verbatim}
\usepackage[english]{babel}
\usepackage{natbib}
\usepackage{ulem}
\usepackage{physics}
\usepackage{booktabs}
\usepackage{enumitem}
\usepackage{mathtools}
\usepackage{url}
\usepackage{multirow}
\usepackage{subcaption}
\usepackage{float}
\usepackage{numprint}
\usepackage{enumitem}
\usepackage[title]{appendix}
\usepackage[font=normalsize, labelfont=bf, justification=centering]{caption}
\usepackage{tabularx}
\usepackage{actuarialangle}
\usepackage{ltablex}
\usepackage{makecell}
\usepackage[hidelinks]{hyperref}
\usepackage{bbm}
\usepackage[table,xcdraw]{xcolor}
\usepackage[ruled,linesnumbered]{algorithm2e}
\makeatletter
\def\ps@pprintTitle{
  \let\@oddhead\@empty
  \let\@evenhead\@empty
  \def\@oddfoot{\centerline{\thepage}}
  \def\@evenfoot{\thepage\hfill}}
\makeatother

\numberwithin{table}{section}
\numberwithin{figure}{section}

\newcolumntype{R}{>{\raggedleft\arraybackslash}X} 
\newcolumntype{P}[1]{>{\centering\arraybackslash}p{#1}} 

\renewcommand\appendix{\par
\setcounter{section}{0}
\setcounter{subsection}{0}
\setcounter{table}{0}
\setcounter{figure}{0}
\gdef\thetable{\Alph{table}}
\gdef\thefigure{\Alph{figure}}
\gdef\thesection{\Alph{section}}
\setcounter{section}{0}}

\numberwithin{equation}{section}

\newcounter{arclist}

\newcounter{arcenum}

\newcommand{\rev}[1]{\textcolor{black}{#1}}

\begin{document}

\normalem

\begin{frontmatter}

\title{Optimal Strategies for the Decumulation of Retirement
Savings \\ under Differing Appetites for Liquidity and Investment Risks}

\cortext[cor]{Corresponding author. }
\address[UMelb]{Centre for Actuarial Studies, Department of Economics, University of Melbourne VIC 3010, Australia}

\author[UMelb]{Benjamin Avanzi}
\ead{b.avanzi@unimelb.edu.au}

\author[UMelb]{Lewis De Felice\corref{cor}}
\ead{lewisdefelice@optusnet.com.au}

\begin{abstract}
A retiree's appetite for risk is a common input into the lifetime utility models that are traditionally used to find optimal strategies for the decumulation of retirement savings.

\rev{In this work, we consider a retiree with potentially differing appetites for the key financial risks of decumulation: liquidity risk and investment risk. We set out to determine whether these differing risk appetites have a significant impact on the retiree's optimal choice of decumulation strategy.} To do so, we design and implement a framework which selects the optimal decumulation strategy from a general set of admissible strategies in line with a retiree’s goals, and under differing appetites for the key risks of decumulation.

Overall, we find significant evidence to suggest that a retiree's differing appetites for different decumulation risks will impact their optimal choice of strategy at retirement. Through an illustrative example calibrated to the Australian context, we find results which are consistent with actual behaviours in this jurisdiction (in particular, a shallow market for annuities), which lends support to our framework and may provide some new insight into the so-called annuity puzzle.

\end{abstract}

\end{frontmatter}

\section{Introduction}\label{sec:intro}
\subsection{Background} \label{ssec:background}
Almost all retirees face the \textit{decumulation problem} of deciding how to best spread their hard-earned savings over the rest of their lives. In general, the retiree seeks to convert their savings into a desired standard of living (SOL) over retirement by choosing a strategy under a complex web of preferences, risks and constraints. Nobel laureate William Sharpe famously coined decumulation to be the \textit{``nastiest, hardest problem in finance''} \citep{Sh17}, attributing the difficulty to the sheer dimensionality of the problem. Furthermore, this complex problem is not limited to the realm of finance alone, placing intergenerational strain on structures of social security and familial support.

An approach to solving the decumulation problem may be characterised by the following three attributes:

\begin{enumerate}
    \item \textit{Exogenous Parameters}. Any solution will be impacted by the retiree's personal circumstances such as age $x$, health status and initial wealth $W_0$ at time of retirement $t=0$. There is also ever-changing public policy such as compulsory minimum rates of drawdown and age pension eligibility criteria.
    \item \textit{Optimisation Framework}. An architecture designed to evaluate a strategy against a stylised interpretation of the retiree's goals and appetite for risk. Although the overarching goal of retirement is late-life fulfillment, goals are typically assumed to be strictly financial. The optimisation framework of choice is traditionally the lifetime utility framework pioneered by \citet{Ya65} and \citet{Me69}. 
    \item \textit{Admissible Strategies}. There is a rich literature which explores different decumulation strategies (and products), with \citet{Pi16}, \citet{PiTa22} and \citet{Pi05} exploring innovative guarantee structures and participation mechanisms. Many of these strategies may help to protect retirement savings as improvement in healthcare continues to extend life expectancy.
    \end{enumerate}

\citet{IFA22} point out the opportunity to extend the set of existing admissible strategies to better allow the retiree to purchase longevity protection as they age, as their priorities shift away from flexibility and efficiency of investment payoff. We remark that this changing demand may indicate the presence of differing appetites for the different risks of decumulation. At any given time, the retiree is unlikely to perceive the risk of experiencing poor investment returns in the same way as the financial risk of living longer than expected. 

Accordingly, \citet{MiHu11} distinguish between a retiree's aversion to `financial' and `longevity' risks, arriving at the natural conclusion that the retiree will consume more conservatively when they are more averse to the risk of living longer than expected. However, the authors form their conclusions from the traditional lifetime utility framework under the simplifying assumption that the retiree holds a risk-free portfolio, allowing them to account for longevity risk aversion through the parameter $\gamma$ of constant relative risk aversion. 

The retirement literature is rich with work which seeks to extend the set of admissible strategies, or determine how the optimal strategy varies across retirees of different personal circumstances. Instead, \citet{MiHu11} effectively varies the optimisation framework to determine how the optimal strategy will change for given exogenous parameters over a narrow set of admissible strategies. We are similarly interested in observing the sensitivity of the optimal strategy to changes in the retiree's risk appetite, but now allowing for a more expansive set of admissible strategies and differing appetites for different risks. This interest also motivates us to develop an optimisation framework which accommodates for the additional complexity. 

If these differing risk appetites \rev{have a} significant \rev{impact on the retiree's optimal choice of decumulation strategy}, there emerges the opportunity to develop attractive \rev{products} which both accommodate for these preferences, and promote the sustainability of our tax and transfer system for generations to come.

\subsection{Contributions}
To the best of our knowledge, there has been no prior literature which has explicitly considered the impact of varying the retiree's differing appetites for different risks on the optimal strategy. We therefore lay the foundations for future work through:
\begin{enumerate}
    \item designing a framework which selects the optimal decumulation strategy from a general set of admissible strategies in line with a retiree's goals and differing appetites for some key risks of decumulation;
    \item finding an admissible set of annuity-like strategies which can be integrated with drawdown to allow us to vary the distribution of outcomes according to the retiree's preferences; 
    \item developing a computationally efficient approach to implement the optimisation framework which conveys how the optimal strategy may vary from retiree to retiree.
\end{enumerate}
Overall, we find significant evidence to suggest that a retiree's differing appetites for different decumulation risks will impact their choice of strategy at retirement. We calibrate our model to the Australian context, and find results which are consistent with the shallow market for annuities observed in this jurisdiction. We discuss how this lends support to our framework, and can provide new insight into the annuity puzzle. 

\subsection{Outline of Paper}
\indent In Section \ref{ch:2}, we discuss two key risks to the retiree: investment risk and liquidity risk, building upon the work of \citet{Pi16} and \citet{Pi07}. The latter refers to the prominent risk associated with the retiree's need for liquidity. In Section \ref{ch:3}, we construct an optimisation framework to select the strategy which best tailors the distribution of outcomes to the retiree's differing risk appetites. We draw on the stochastic present value (SPV) methodology of \citet{Mi97} to determine the distribution of outcomes to the retiree. We follow a similar approach to \citet{Ol01} to decompose the distribution of outcomes based on investment risk and liquidity risk, and find the optimal strategy based on the retiree's differing appetites for these risks. In Section \ref{sec:integ}, we find a set of admissible strategies in a generalised family of annuities which we construct with inspiration from \citet{Pi05}. In Section \ref{ch:6}, we develop an approach to implement our optimisation framework in a general context, balancing computational efficiency and numerical precision. We then apply our implementation approach in Section \ref{sec:aus_con} through the illustrative example of the Australian context. 

\section{Outline of the Key Decumulation Risks}\label{ch:2}
We define a \textit{decumulation risk} as a source of uncertainty with bidirectional (both upside and downside) impact on the retiree's goals. We focus on  financial goals which surround consumption and bequest, as discussed in Section \ref{sec:plan}.

\subsection{Liquidity Risk}\label{sec:trans}
We attribute \textit{liquidity risk} to the uncertain timing of all liquidity events for which there is demand for significant withdrawal of retirement savings, permanently changing the retiree's financial circumstances. We emphasise that liquidity risk cannot be well-diversified since the retiree only lives out their retirement once.

Liquidity risk can stem from \textit{death risk} due to the distribution of the random remaining time $T_x$ until the event of death; and \textit{long-term care} (LTC) \textit{risk} due to the random time $T_x^{\mathrm{LTC}}$ until the event of entering permanent care. Furthermore, liquidity risk can arise from a number of events such as theft, natural disaster or the need to help children for a house deposit prior to death. In general, we define $T_x^{\mathrm{ELN}}$ as the time of first early liquidity need prior to death (if any), such that $T_x^{\mathrm{ELN}}\leq T_x^{\mathrm{LTC}}$.

We define the \textit{first transition time} by $T_x^* = \mathrm{min}(T_x, T_x^{\mathrm{ELN}})$, reflecting the time at which the fund balance must be significantly withdrawn upon and circumstances permanently change. We consider these random transition times on support $\mathcal{S}$ where $\bar{T} = \sup\mathcal{S}$ is the final possible time of transition. 

\subsubsection{Death Risk} Referring to \citet{Wa10}, we attribute death risk to the uncertainty surrounding the length of the retiree's own remaining lifetime $T_x$. We distinguish between \textit{late} death risk from $T_x$ realising a relatively large value and \textit{early} death risk from $T_x$ realising a relatively small value. A retiree is at risk of outliving their wealth in event of late death, or leaving an incidentally large bequest in event of early death. 

Whether a death time is considered early or late will depend on the objective function under consideration, with life expectancy $\mathbb{E}[T_x]$ \rev{not always serving the most appropriate benchmark. For instance, the expected value of a non-linear function in $T_x$ (e.g. life annuity) is unlikely to be a simple function of $\mathbb{E}[T_x]$.} When a retiree participates in a pool of $n$ lives with remaining lifetimes $\{T_{x; i}:i = 1,\dotsc, n\}$, the earliness of the retiree's death time can be assessed relative to the mortality experience of these pooled lives. Following \citet{Zh20}, we can quantify the mortality experience of the pool through observed survival probability of the pool $_tP_x^{(n)}$:
\begin{equation}\label{eq:conm}
_tP_x^{(n)} = \dfrac{1}{n}\sum_{i=1}^n \mathbbm{1}_{\{T_{x;i} > t\}}. \end{equation}

The retiree is therefore exposed to the \textit{aggregate death risk} which arises from uncertainty in $_tP_x^{(n)}$ relative to projected survival probabilities $_tp_x = \mathbb{E}[\mathbbm{1}_{\{ T_x > t\}}]$. We refer to the terminology of \citet{Pi07} to decompose aggregate death risk into idiosyncratic and systematic components.

\citet{Pi07} defines \textit{mortality risk} as the risk of deviation in the survival of the pool $_tP_x^{(n)}$ from projected survival probabilities $_tp_x$ due to idiosyncratic variation amongst $\{T_{x; i}\}$. As per \citet{BoFrRu23}, idiosyncratic mortality fluctuations in portfolios of limited size can be diversified in large populations. By the Law of Large Numbers, it follows that $_tP_x^{(n)}\to  \,_tp_x$ for $\{T_{x; i}\}$ independent and identically distributed. However, $\{T_{x; i}\}$ are not independent and $n$ is far from infinite across nations with shallow annuity markets.

\citet{Pi07} defines \textit{longevity risk} as the risk of deviation in the observed survival of the pool $_tP_x^{(n)}$ from the projected survival probabilities $_tp_x$ due to systematic variation in population longevity from projections. We define $_tp_x^{\mathrm{ref}}$ as the survival probability associated with a given reference population selected at inception of the pooling policy. \citet{Ca08} claim that longevity risk is non-diversifiable since it affects all individuals, stemming from uncertainty in forecasting the distribution of population rates $_tp_x^{\mathrm{ref}}$. Where pooled lifetimes $\{T_{x; i}\}$ are all sampled with certainty from this reference population, we can define $_tp_x^{\mathrm{ref}} = \lim_{n\to \infty} \,_tP_x^{(n)}$ since we will be pooling over the population in the limit.

\citet{Bl18} recognises longevity risk as a dangerous `trend risk', with \citet{Li13} claiming it poses a gradually developed `chronic' problem to annuity providers. Where the projected death hazard rate $\lambda_{x+t}$ happens to systematically understate or overstate the observed hazard rate $\Lambda_{x+t}$ of $ _tp_x^{\mathrm{ref}}$, the deviation of $ _tp_x^{\mathrm{ref}}$ from $ _tp_x$ will structurally increase over time rather than be cyclical or self-correcting. Due to this `toxicity' of longevity risk, \citet{EvSh10} suggest there is limited appetite in the market to reinsure this risk, forcing an annuity provider to impose higher risk margins in pricing.

\subsubsection{LTC Risk} 
We attribute LTC risk to the randomness of timing $T_x^{\mathrm{LTC}}$ in when significant unplanned costs are incurred prior to death (if at all), due to the onset of senescent disability and permanent need for support \citep{Pi16}. \citet{Ri14} argues that retirement savings should be invested for the different `phases' of retirement as needs change, so we might consider $T_x^{\mathrm{LTC}}$ as the transition time (if any) from the \textit{active phase} to the \textit{frail phase} of retirement. The risk of requiring LTC prior to death has become especially prominent due to the dangerous trend of rising LTC costs in excess of general price inflation \citep{RaOg18}. 

There is a potentially perilous interaction between death risk and liquidity risk. According to \citet{Sh08}, there is an intuitive negative correlation between the liquidity risk and late death risk since disability is more likely to culminate in early death. However, \citet{Zh11} and \citet{Pi16} remark that later death times can lengthen the duration of the frail phase prior to death, incurring more LTC costs and reflecting a possible form of positive correlation between the two risks. 

 Following \citet{Pi16}, 
it is natural to consider the phase $T_x - T_x^{\mathrm{LTC}}$ as more likely to be longer in duration when $T_x$ is large, conditional on the retiree incurring LTC needs prior to death.

\subsection{Investment Risk}\label{sec:inv_unc}
We attribute \textit{investment risk} to randomness in the path of \textit{real} portfolio returns, as impacted by nominal portfolio returns and price inflation. In a decumulation context, we consider the risky cost of funding a unit cash flow at future time $t$ in today's dollars. We denote this cost by discount factor $I(t; \mathbf{u})$ for $t\in \mathcal{S}$ and $\mathbf{u}$ is the trading strategy driving the retiree's investment account $A(\mathbf{u})$. We can further write $I(t; \mathbf{u})= Y(t; \mathbf{u})\Pi(t)$ where the random portfolio discount factor $Y(t; \mathbf{u})$ is the source of \textit{portfolio risk} and the random price level $\Pi(t)$ is the source of \textit{inflation risk}. In particular, we might regard $\Pi(t)$ as the price level associated with the retiree's own unique basket of goods and services.

\subsubsection{Portfolio Risk} We consider portfolio risk to surround the sequence and scale of market returns due to randomness in the nominal rate of return on investment. A retiree will be averse to a sequence of either volatile or unexpectedly high discount factors $Y(t;\mathbf{u})$, creating uncertainty around the amount of investment today required to fund their future cash flows. 

In its most basic form, portfolio risk is often discussed in the context of Modern Portfolio Theory \citep[MPT, originally introduced by][]{Ma52}, typically when considering an investment return over a single period. Unlike many traditional investments, portfolio uncertainty in decumulation must be managed over a sequence of periods. In fact, \citet{FrBl10} found that uncertainty in the sequence of returns can have a greater impact on retirement outcomes than the total rate of portfolio growth over retirement. Retiring into an unexpected bear market can be more dangerous than experiencing the same poor returns closer to the end of one's lifetime, since stable withdrawals will leave fewer assets behind to benefit from a longer future of growth. In a decumulation context, it is therefore vital not to assess the performance of a portfolio solely by the rate of return over a time horizon without regard for consumption over that same horizon \citep{BeDo18}. 

Furthermore, \citet{Re95} discusses the phenomenon of `time diversification' where portfolios become less volatile in value over longer time horizons. As such, \citet{FrMiBl10} find shorter periods of decumulation to be more sensitive to adverse sequences of market returns since there exists less time for the market to recover, presenting an even greater danger to older retirees who plan over shorter time horizons. 

\subsubsection{Inflation Risk} We attribute inflation risk to the randomness in the purchasing power of each dollar of the retiree's savings over time. This uncertainty arises largely from randomness in the price inflation of the retiree's everyday basket of goods and services as reflected by price level $\Pi(t)$. The potential for variability in $\Pi(t)$ may create further uncertainty around the amount of investment today required to fund a given future cash flow. Moreover, the relationship between portfolio returns and inflation has tended to vary uncertainly over time, depending on factors such as the function of monetary policy in the economy \citep{Zh21}.

Inflation risk is exacerbated in the decumulation phase, with little ability to hedge the impact of high inflation through rising wages \citep{Fu08}. This exposure extends over many decades, especially eroding the purchasing power of those individuals invested in less risky assets \citep{DoKhLi23} with little potential to receive high long-term rates of portfolio return. Though inflation-indexed bonds present themselves as an attractive hedge, the market for them is relatively illiquid in practice \citep{Fi18} and the investor will be forced to accept lower yields in order to transfer the risk. Though, we remark that the purchasing power required to sustain a given standard of living will gradually decline as the retiree ages and their everyday spending needs diminish \citep{Hu08}, reflecting a form of hedge against late-life inflation risk.

\section{Optimisation Framework}\label{ch:3}
We construct a framework to find the decumulation strategy which best tailors the distribution of outcomes to the retiree's differing risk appetites. 

\subsection{Aligning the Retiree's Goals with a Retirement Plan}\label{sec:plan}
\label{sec:obj} 
Although the overarching goal of retirement is late-life fulfillment, we will focus on the pursuit of a set of strictly financial goals set at retirement time $t=0$ as below:
\begin{enumerate}[label = (\roman*)]
    \item Have the ability to sustain a chosen standard of living (SOL) over time with flexibility to adjust spending as early liquidity needs are incurred over time;
    \item Have adequate savings upon death in line with some general bequest motive (if any).
\end{enumerate}
\citet{Pf15} calls on the value of actuarial science in modelling a \textit{consumption target} in the style of a defined benefit pension. \citet{Yi23} similarly sets consumption to achieve a constant pre-retirement SOL in order to 
benchmark the adequacy of retirement savings. To evaluate decumulation strategies, we therefore define the following \textit{retirement plan}:
\begin{enumerate}
\item  We initially fix the consumption target $\left(c(t)\right)_{t\in \mathcal{S}}$ in real terms to best satisfy Goal (i) with the intention of changing this consumption target as early liquidity needs are incurred over time;
\item The cost of funding $\left(c(t)\right)_{t\in \mathcal{S}}$ is planned to deliver a bequest in line with Goal (ii), provided the initial consumption target can be maintained until death.
\end{enumerate} 

In pursuit of the retirement plan, the retiree will face the investment risk of having to reduce their bequest upon realisation of poor investment returns. They also face the liquidity risk of having to reduce their bequest due to unfavourable death time $T_x$; or having to bring forward their intended bequest upon realisation of event $T_x^{\text{ELN}} < T_x$. Indeed, \citet{Lo14} describes how intended bequests can function as an incidental form of self-insurance against large costs associated with LTC needs, which we generalise to allow for all early liquidity needs. We will observe these risks emerge in Section \ref{sec:em_risk}.

At time $T_x^*$, consumption target $\left(c(t)\right)_{t\in \mathcal{S}}$ is considered to become no longer viable as circumstances significantly change. The retiree withdraws remaining funds for immediate bequest where $T_x^* = T_x$; or the retiree has the option to recast their retirement plan upon $T_x^* = T_x^{\mathrm{ELN}}$.

\subsection{Distribution of Outcomes under Decumulation Strategies}\label{sec:deccost}
 We define a decumulation strategy as any mechanism through which the \rev{retiree} can convert their wealth into the retirement plan in Section \ref{sec:obj}. We introduce the \textit{residual liquidity function} $X_{\angl{T}}$ to track the retiree's outcome by time $T$ against the retirement plan. We define $X_{\angl{T}}$ as some measure of wealth available to the retiree at time $T$, reflecting the retiree's remaining budget for future consumption, early liquidity needs and bequest. This budget will be random due to the risks in Section \ref{ch:2}, giving rise to a distribution of possible outcomes. Where $T = T_x^{\text{ELN}}$, we may interpret $X_{\angl{T}}$ as the budget for the frail phase that remains after expenditure over the active phase, allowing the retiree to recast their retirement plan and decumulation strategy as their needs change.

\citet{MiRo05} apply a stochastic present value (SPV) approach in their analytical approach in deriving a sustainable consumption rate $c(t)\equiv c$, with \citet{Di09} arguing that the SPV is a valuable tool for comprehensive actuarial risk management of funding commitments. We evaluate $X_{\angl{T}}$ by the SPV of $W_T$ as a proportion of $W_0$ in order to represent the portion of each dollar invested upon retirement which remains liquid at time $T$. We use $W_{T}$ and  $W_T^*$ to denote the retiree's nominal and real wealth, respectively, and give two equivalent expressions for $X_{\angl{T}}$:\begin{equation}\label{eq:spv}
X_{\angl{T}} = Y(T; \mathbf{u}) \frac{W_{T}}{W_0} = I(T; \mathbf{u}) \dfrac{W_T^*}{W_0}.   
\end{equation}
In the following subsections, we find an expression for $X_{\angl{T}}$ under a pure drawdown strategy; and use this expression to shed light on the decumulation risks associated with this strategy.

\subsubsection{Distribution of Outcomes under a Pure Drawdown Strategy} We first consider the standard strategy of \textit{pure drawdown} from investment account $A(\mathbf{u})$ for comparison against alternative strategies.
 Assuming no inflows from family, social security or corporate pensions, a pure drawdown strategy will decompose each retirement dollar as follows:
\begin{equation}\label{eq:dd}
1 = X_{\angl{T}} + \int_0^T I(t; \mathbf{u})c(t)\dd{t}. \end{equation}
While (\ref{eq:dd}) is stated in proportions of $W_0$, we can informally interpret (\ref{eq:dd}) as describing the decumulation of $\$1$ over $[0, T]$ to fund consumption target $(c(t))_{t\in \mathcal{S}}$, with $X_{\angl{T}}$ remaining at time $T$. The integral in (\ref{eq:dd}) can be retrieved from \citet{Mi97} in the special case of a `stochastic perpetuity' as $T\to \infty$ under assumptions of constant nominal consumption $c(t) = c$ and stochastic Wiener returns.

\subsubsection{Understanding Emergence of Risk under Pure Drawdown}\label{sec:em_risk}
We now seek to understand the risks which shape the distribution of outcomes. To do so, we may compare the observed outcome $X_{\angl{T}}$ by time $T$
with the \textit{planned bequest} $\mathbb{E}[X_{\angl{T_x}}]$. The planned bequest will be expected upon retirement, conditional on consumption target $(c(t))_{t\in\mathcal{S}}$ being maintained until death.

Since the integrand of $1 - X_{\angl{T_x}}$ is non-negative and measurable, we can apply Tonelli's Theorem to interchange the order of integration when considering planned bequest $\mathbb{E}[X_{\angl{T_x}}]$, giving deterministic integrand $z(t) = c(t)\mathbb{E}[I(t; \mathbf{u})\mathbbm{1}_{\{T_x > t\}}]$. The retiree therefore plans their initial wealth (or \$1) to be decumulated over $\mathcal{S}$ with $\bar{T} = \sup \mathcal{S}$ as follows:

\begin{equation}\label{eq:expe}
1 = \mathbb{E}[X_{\angl{T_x}}] + \int_0^{\bar{T}} z(t) \dd{t}.
\end{equation}
For $T$ and $(I(t;\mathbf{u}))_{t\in\mathcal{S}}$ independent, we rewrite (\ref{eq:dd}) as follows:
\begin{equation}\label{eq:dec}
1 = X_{\angl{T}} +  \int_0^{\bar{T}} z(t) K(t; \mathbf{u})\dd{t},\end{equation}where
\begin{equation}\label{eq:K}
K(t; \mathbf{u}) = \dfrac{I(t; \mathbf{u})}{\mathbb{E}\left[I(t; \mathbf{u})\right]}\cdot\dfrac{\mathbbm{1}_{\left\{T > t\right\}}}{_tp_x}. \end{equation}

We observe in (\ref{eq:K}) that function $K(t; \mathbf{u})$ measures the random variation in the retiree's experience from the retirement plan over each interval $[0, t]$. By setting $T = T_x^*$, we recognise these deviations as arising from the ensemble of risks outlined in Chapter \ref{ch:2}. When $I(t;\mathbf{u})$ is generally high or $T$ is large in (\ref{eq:K}), then the integral of (\ref{eq:dec}) will end up occupying a greater share of the retiree's initial wealth (or \$1) than otherwise. That is, pure drawdown can be vulnerable to investment risk and late death risk. Conversely, the integral will shrink when $T$ is small, as reflective of flexible access to funds in event of liquidity need.

\subsubsection{Combining Drawdown with Alternative Strategies}
We now consider combining drawdown with some general alternative strategy. In Section \ref{sec:integ}, we will consider pooling policies (e.g. annuities) as a mechanism to mitigate the risks associated with pure drawdown.

In discussion of annuities, \citet{Ya65} introduces the notion of a `flow of earnings' into the retiree's budget constraint. To reinforce that these `earnings' stem from employment over the accumulation phase prior to time $t = 0$, we instead define a random \textit{payment stream} $(d(t))_{t\in \mathcal{S}}$ for which the retiree pays \textit{premium} $\$ P_{\theta}$. We consider each payment $d(t)$ to flow from \textit{alternative account} $A(\tilde{\mathbf{u}})$ with trading strategy $\tilde{\mathbf{u}}$ into the retiree's investment account $A(\mathbf{u})$ for the purpose of drawdown. For consistency with $c(t)$, we state each payment $d(t)$ and premium $P_{\theta}$ at time $0$ in real terms as proportions of $W_0$. Given the retiree now has $\$(1 - P_{\theta})$ to decumulate over $[0, T]$, (\ref{eq:dd}) generalises as follows to give an expression for the \textit{overall decumulation strategy}:
\begin{equation}\label{eq:dd1}
1 - P_{\theta} = X_{\angl{T}} + \int_0^T I(t; \mathbf{u})\big(c(t) - d(t)\big)\dd{t}. \end{equation}
As per (\ref{eq:dd1}), we can consider a retiree's overall decumulation strategy to be composed of the investment account $A(\mathbf{u})$ and payment stream $(d(t))_{t\in \mathcal{S}}$. Following the suggestions of \citet{BeDo18} and \citet{IFA22}, this construction also gives the retiree the autonomy to convert payment $d(t)$ into consumption or bequest at some future time $s\geq t$ based on their needs. 

\subsection{Decomposing the Distribution of Outcomes by Decumulation Risk}\label{sec:rd}
We seek to decompose the distribution of outcomes in order to apply the retiree's differing risk appetites for liquidity risk and investment risk. For motivation, we look to \citet{Ol01} who separates longevity risk from the \textit{total cost function} $C_x^{(n)}$ associated with a provider's annuity portfolio of $n$ lifetimes $\{T_{x;i}\}$. They decompose the variance of $C_x^{(n)}$ conditional on observed population survival $\mathcal{P} = (\,\!_tp_x^{\mathrm{ref}})_{t\in \mathcal{S}}$ as follows:
\begin{equation}\label{eq:var}
\mathrm{Var}\big(C_x^{(n)}\big) = \mathbb{E}\big[ \mathrm{Var}(C_x^{(n)} \vert \, \mathcal{P})\big] + \mathrm{Var}\big( \mathbb{E}[C_x^{(n)} \vert \, \mathcal{P}]\big). \end{equation}
As described by \citet{Wa10}, the first term in (\ref{eq:var}) reflects the risk to the annuity portfolio if there was no longevity risk, with the second term reflecting the effect due to longevity risk. We apply a similar risk decomposition to $X_{\angl{T}}$ in order to adopt the perspective of the retiree, but without necessarily assuming that variance is the best measure of each risk. We consider the impact of each decumulation risk through constructing the following two information sets:
    \begin{enumerate}
        \item $\mathcal{L}$: information set where transition times are random, with all else occurring as planned. We define $X_x^{\mathcal{L}} = \mathbb{E}[X_{\angl{T}}\vert \mathcal{L}]$ for $T = T_x^*$ in order to consider the impact of uncertainty in the first transition time on the liquidity available at that time. Taking conditional expectations of \eqref{eq:dec}, we find
\begin{equation}\label{eq:K1}
        1 = X_x^{\mathcal{L}}+ \int_0^{\bar{T}} z(t)\cdot \dfrac{\mathbbm{1}_{\{T_x^* > t \}}}{_tp_x}\dd{t}
        \end{equation}
        When $T_x^*$ takes larger values, we observe in (\ref{eq:K1}) that the liquidity $X_x^{\mathcal{L}}$ available to the retiree at $T_x^*$ will fall as a share of their initial wealth (or \$1).
        We have removed the investment risk from function $K(t;\mathbf{u})$ in (\ref{eq:K}), allowing us to analyse liquidity risk independently of investment risk.     
        \item $\mathcal{I}$\,: information set where investment returns are random, with all else occuring as planned. We define $X_x^{\mathcal{I}} = \mathbb{E}[X_{\angl{T}}\vert \mathcal{I}]$ for $T = T_x$ in order to consider the impact of uncertainty in investment returns on bequest, provided there are no early liquidity events. Taking conditional expectations of (\ref{eq:dec}), we find
\begin{equation}\label{eq:K2}
        1 = X_x^{\mathcal{I}}+ \int_0^{\bar{T}} z(t)\cdot \dfrac{I(t;\mathbf{u})}{\mathbb{E}[I(t;\mathbf{u})]}\dd{t}
        \end{equation}
        When $I(t;\mathbf{u})$ is generally higher than $\mathbb{E}[I(t;\mathbf{u})]$, we observe in (\ref{eq:K2}) that the bequest available to a retiree at $T_x$ will fall as a share of their initial wealth (or $\$1$). We have removed the liquidity risk from function $K(t;\mathbf{u})$ in (\ref{eq:K}), allowing us to analyse investment risk independently of liquidity risk.
    \end{enumerate}

From (\ref{eq:K1}) and (\ref{eq:K2}), it follows that information sets $\mathcal{L}$ and $\mathcal{I}$ will generate two independent distributions of outcomes. In Section \ref{ch:5}, we will seek out the optimal decumulation strategy according to the retiree's appetites for variability across each of these two distributions, thereby optimising the overall distribution of outcomes.

\subsection{Optimising the Distribution of Outcomes for Varying Risk Appetites}\label{ch:5}
With reference to the traditional lifetime utility approach, we will now develop risk measures which capture the unique dangers of each independent distribution, and combine them into an optimisation problem to be solved.

\subsubsection{Brief Review of Lifetime Utility Optimisation} 
The problem of finding the optimal decumulation strategy which best satisfies Goal (i) and Goal (ii) is traditionally solved over planning horizon $\mathcal{S} = [0, \bar{T}]$ through maximising expected utility of lifetime consumption $(c(t))_{t\in \mathcal{S}}$ and bequest $W_T$ at terminal time $T\leq \bar{T}$. \citet{Ya65} pioneered the problem of best decumulating retirement savings in a setting where $T \equiv T_x$, and the retiree faces only death risk. \citet{Me69} introduces investment risk into the problem for fixed death time $T_x\equiv t^*$, putting forward the Classical Portfolio Problem of jointly finding trading strategy $\mathbf{u}$ and $(c(t))_{t\in \mathcal{S}}$ to maximise expected lifetime utility. Given time-preference parameter $\rho$ and general bequest utility function $u_B(\cdot)$, the objective function can be expressed as follows:
\begin{equation}\label{eq:Q}
Q_{\angl{T}}  = \mathbb{E}\left[\int_0^{T} \mathrm{e}^{-\rho t} u(c(t))\dd{t} + u_B(W_{T}, T) \right]. \end{equation}
We leverage the \textit{traditional approach} of maximising lifetime utility to optimise the distribution of outcomes over the $\mathcal{I}$ information set. We also discuss the shortfalls of using this traditional approach to reflect optimality over the $\mathcal{L}$ information set.
\subsubsection{Construction of Investment Objective Function} 
 Where consumption target $(c(t))_{t\in\mathcal{S}}$ is predetermined according to some chosen standard of living, the problem of maximising (\ref{eq:Q}) reduces into the problem of maximising $u_B(W_T, T)$. \citet{Lo12} elects to maximise the utility of the present value of bequest, remarking that the longer one lives, the `cheaper' it is to leave a bequest of given present value. We therefore consider the retiree to maximise their expected utility of bequest $Q_x^{\mathcal{I}} = \mathbb{E}[u(X^{\mathcal{I}}_x)]$ in the $\mathcal{I}$ information set. 
 
 In our choice of objective function, we take motivation from \citet{La18} who find a dynamic trading strategy $\mathbf{u}$ through mean-variance optimisation of bequest given a level of risk aversion that depends on the state of the investment. In the spirit of MPT, we can consider the retiree to be a mean-variance investor with respect to $X^{\mathcal{I}}_x$ to give the \textit{investment objective function}: 
\begin{equation}\label{eq:objj}
 Q_x^{\mathcal{I}} = \mathbb{E}[X_{x}^{\mathcal{I}}] - b \mathbb{E}[(X_{x}^{\mathcal{I}})^2],
 \end{equation}
 where we define $b \geq 0$ as the parameter of \textit{investment risk aversion} (IRA). For values of $X_x^{\mathcal{I}}$ where $u(X_x^{\mathcal{I}})$ is concave, we assume the retiree derives higher utility from a lower funding cost of consumption; and being able to reserve a greater proportion of their initial wealth for bequest. However, \citet{BeDo18} deem it `wasteful' to pay for surpluses of assets in excess of a target. In our setting, the bliss point $X_x^{\mathcal{I}} = 1/2b$ may therefore be interpreted as the maximum share of initial wealth that the retiree may wish to bequeath. 

\subsubsection{Construction of Liquidity Risk Constraint}\label{sec:liq}
While the traditional approach can yield tractable solutions, it can be limited in its ability to convey the retiree's appetite for the liquidity risk associated with random time $T_x^*$. 

In particular, \citet{BoVi12} argue that the traditional approach is constrained by an assumed form of risk neutrality in $Q_{\angl{T}}$ with respect to random death time $T = T_x$. This risk neutrality is said to arise from the central assumption of (\ref{eq:Q}) that consumption utility is \textit{additively separable}, that is integrable over time, originally dubbed an `unhappy' assumption by \citet{Ya65}.

Consequentially, \cite{Da05} maximise lifetime utility to deduce that retirees derive higher utility from laying aside a fixed sum at a fixed time for bequest before annuitising their remaining wealth. This is because bequest $W_{T_x}$ will be random in size and timing where $T_x$ is uncertain, illustrating the failure of the traditional approach to capture a retiree's lower tolerance to leaving a smaller bequest in the event of very early death.

Given these limitations, we take an alternative approach to incorporating liquidity risk appetite. \citet{BoFa23} propose a penalised utility function in the context of portfolio theory, tilting the optimal portfolio based on the manager’s desires for exposure to different risk factors. \citet{BaYo12} propose a similar penalty in the context of the traditional approach, maximising the lifetime utility of consumption subject to a binding constraint on the probability of ruin. 

We similarly maximise $Q_x^{\mathcal{I}}$ in (\ref{eq:objj}), subject to a maximum constraint on the probability of shortfall in the $\mathcal{L}$ information set. We allow the retiree to select the degree of certainty with which they want to have a given amount of liquidity in time of need, assuming investments perform as planned. We define $\tau_{\nu}$ as the time of shortfall in $X_x^{\mathcal{L}}$ relative to \textit{shortfall threshold} $\nu$. We impose a \textit{liquidity risk constraint} upon the \textit{liquidity shortfall probability} $\mathbb{P}_{\mathcal{L}}(T_x^* \geq \tau_\nu)$ associated with falling below $\nu$ by time $T_x^*$:
 \begin{equation}\label{eq:con} \mathbb{P}_{\mathcal{L}}(T_x^* \geq \tau_\nu ) \leq \psi_{\nu}, \end{equation}
where we define $\psi_{\nu}$ as the parameter of \textit{liquidity risk tolerance} (LRT) associated with shortfall threshold $\nu$. Though we will assume a single value of $\nu$ for each retiree, their liquidity risk appetite might be better parameterised by some set of tolerances $\{\psi_{\nu}: \nu \in \mathcal{A}\}$ associated with threshold set $\mathcal{A}$, or even allowing these tolerance(s) to vary across different sources of liquidity risk. Moreover, \citet{Po95} takes the Laplace Transform of the (random) ruin time as a way of conveying the higher cost of relatively early insolvency to an insurer. Since $T_x^* \leq T_{x}$, we are similarly assigning a higher penalty to the more costly possibility of early shortfall.  

\subsubsection{Summary of Optimisation Problem}\label{sec:final_opt}
In summary, we seek the optimal admissible strategy which achieves the following maximisation:
\begin{equation}
Q_x^{\mathcal{I}} = \mathbb{E}[X_x^{\mathcal{I}}] - b\mathbb{E}[(X_x^{\mathcal{I}})^2] \ \quad  \text{subject to} \quad \ \mathbb{P}_{\mathcal{L}}(T_x^* \geq \tau_{\nu}) \leq \psi_{\nu}. \end{equation}
Overall, we parameterise the retiree's risk appetites by $(b, \psi_{\nu})$ for some single choice of $\nu$. In Section \ref{sec:integ}, we will construct a family of admissible strategies over which to perform this optimisation.

\section{Finding a Set of Admissible Strategies}\label{sec:integ}

In Section \ref{ch:3}, we developed a framework to find the optimal strategy from a general set of $L$ alternative strategies $\{d_i(t): i = 1,\dotsc, L \}$ which may be chosen at retirement in combination with a baseline strategy of drawdown from investment account $A(\mathbf{u})$. We discuss pooling mechanisms in Section \ref{sec:dyn}, allowing us to define a family of alternative strategies in Section \ref{sec:ann_fam}. In Section \ref{sec:dec_v}, we admit this family as the admissible solution set to our optimisation problem under some restrictions.

\subsection{Dynamics of Risk Transfer via Pooling}\label{sec:dyn} We consider \textit{pooling policies} to comprise a stream of payments $(d(t))_{t\in \mathcal{S}}$ receivable until death time $T_x$. The retiree can purchase a mortality credit guarantee (e.g. life annuity) from a \textit{pooling provider}, or participate directly in the experience of the pool through the mutuality mechanism. The amounts released by the deceased are shared as mortality credits amongst the survivors in the participating case \citep{Pi16}, and are otherwise credited to the pooling provider. 

In either case, the actuarially fair price of the pooling policy to the retiree is $P_0 = \mathbb{E}[C_{\angl{T_x}}]$, where we refer to $C_{\angl{T_x}}$ as the \textit{individual cost function} associated with funding $(d(t))_{t\in \mathcal{S}}$. Further, $C_{\angl{T_x}}$ will contain the variability transferred to the pool and can be written:
\begin{equation}\label{eq:3}
C_{\angl{T_x}} = \int_0^{T_x} I(t; \tilde{\mathbf{u}}) d(t) \dd{t}.
\end{equation}
The variability of $C_{\angl{T_x}}$ is passed through the pooling mechanism to diversify away idiosyncratic variability associated with portfolio returns and individual lifetime $T_x$. The pooling provider is left with a portfolio associated with total cost function $C_x^{(n)}$, as introduced in Section \ref{sec:rd}. We write
\begin{equation}\label{eq:tot_cost}
C_x^{(n)} = \sum_{i = 1}^n C_{\angl{T_{x;i}}} = \sum_{i = 1}^n \int_0^{T_{x; i}} I(t; \tilde{\mathbf{u}}) d(t) \dd{t}.
\end{equation}
We rewrite this exposure in terms of the \textit{average cost function} $\bar{C}_x^{(n)}$:
\begin{equation}\label{eq:cost}
\bar{C}_x^{(n)} = \dfrac{1}{n} \sum_{i = 1}^n \int_0^{T_{x; i}} I(t; \tilde{\mathbf{u}}) d(t) \dd{t} = \int_0^{\bar{T}} I(t; \tilde{\mathbf{u}}) d(t) \,_tP_x^{(n)}\dd{t}.\end{equation}

By taking the arithmetic average of individual cost functions $C_{\angl{T_{x;i}}}$, we assume without loss of generality that all members make the same initial investment to the pool. Otherwise, we could redefine $_tP_x^{(n)}$ by some weighted average that accounts for differences in initial investment amount across pooled lives. By definition of $d(t)$, it follows that $\bar{C}_x^{(n)}$ will be stated as a proportion of the retiree's own initial wealth $W_0$.

A pooling provider will bear the variability of $\bar{C}_x^{(n)}$ in return for a higher price $P_{\theta} = (1 + \theta)P_0$, marked up in comparison to the actuarially fair price $P_0$.
A \textit{loading factor} $\theta$ will arise from the purchase of a mortality credit guarantee, reflecting the cost of transferring risk \citep{Pi16}. Indeed, $\bar{C}_x^{(n)}$ reflects the provider's exposure to \textit{asset-liability mismatch risk} which we define as the risk that the value of liabilities moves differently to the value of assets held to back them. In our setting, assets are subject to investment risk due to $I(t; \tilde{\mathbf{u}})$; and liabilities are subject to aggregate death risk due to $_tP_x^{(n)}$.
\subsection{Annuity Family}\label{sec:ann_fam}
In the following subsections, we develop a family of annuity-like products which are to be admissible in our optimisation framework. We showcase the decumulation risks associated with these products, as well as their risk mitigation capabilities.

\subsubsection{Constructing the Annuity Family}\label{sec:gsa_fam}
We develop a family of pooling policies based on the original \textit{group self-annutisation} (GSA) formulation due to \citet{Pi05} and discussed at depth by \citet{Zh20} and \citet{BaGa22}. We denote $d(t) = \phi g(t)$ where $\phi$ is the \textit{payment rate} and $g(t)$ is the \textit{payment vehicle}. We write
\begin{equation}\label{eq:gsa_eq}
g(t) =
\xi(t)\cdot \dfrac{\mathbb{E}[I(t; \tilde{\mathbf{u}})]}{I(t; \tilde{\mathbf{u}})} \cdot \dfrac{_tp_x}{_tP_x^{(n)}}.\end{equation}
\noindent For illustration, we can rewrite (\ref{eq:dd1}) under (\ref{eq:gsa_eq}) as follows:

\begin{equation}\label{eq:pro0}
1 - P_{\theta} = X_{\angl{T}} + \int_0^T I(t; \mathbf{u})\big(c(t) - \phi g(t)\big)\dd{t}. \end{equation}

\citet{Pi05} denotes the factors of $g(t)$ by the interest rate adjustment $\mathrm{IRA}_t = \mathbb{E}[I(t; \tilde{\mathbf{u}})] / I(t; \tilde{\mathbf{u}})$ and mortality experience adjustment $\mathrm{MEA}_t = \,_tp_x/\!\,_tP_x^{(n)}$, respectively. In our inflationary setting, we implicitly incorporate an inflationary adjustment into the $\mathrm{IRA}_t$ factor based on price level $\Pi(t)$. We generate the \textit{Annuity Family} by varying the \textit{guarantee factor} $\xi(t)$ in (\ref{eq:gsa_eq}) to undo these adjustments and transfer the risk to a provider.

Given \textit{mortality credit guarantee} where $\xi(t)\cdot\mathrm{MEA}_t \equiv 1$, we get a \textit{unit-linked annuity} (ULA). Given \textit{portfolio return guarantee} where $\xi(t)\cdot\mathrm{IRA}_t \equiv 1$, we retrieve the \textit{longevity-indexed annuity} (LIA) due to \citet{De11}. Where $\xi(t)\cdot\mathrm{MEA}_t\cdot \mathrm{IRA}_t \equiv 1$, we recover an \textit{inflation-indexed annuity} (IIA) with $d(t) = \phi$. We summarise the Annuity Family in Table \ref{table:taby} below:

\begin{table}[hbt!]
$$
\begin{array}{r||l|l}
\text{Annuity Family Policy} & \text{Nature of Guarantee} & \text{Payment Vehicle} \ g(t)\\ \hline \hline
\text{Group Self-Annuitisation (GSA)}& \text{None} &\mathrm{IRA}_t\cdot\mathrm{MEA}_t\\ \hline
\text{Unit-Linked Annuity (ULA)} & \text{Mortality Credit} & \mathrm{IRA}_t\\ \hline
\text{Longevity-Indexed Annuity (LIA)}& \text{Portfolio Return} & \mathrm{MEA}_t \\ \hline
\text{Inflation-Indexed Annuity (IIA)} & \text{Mortality Credit}  & 1\\
& \text{Portfolio Return} &
\end{array}
$$
\caption{Annuity Family Payment Vehicles}
\label{table:taby}
\end{table}

\subsubsection{Mitigation and Emergence of Risks under Combined Drawdown with Annuity Family}\label{sec:prot}
 We seek to incorporate payment streams $(d(t))_{t\in \mathcal{S}}$ from the Annuity Family into the retiree's overall decumulation strategy to mitigate against the risks of (\ref{eq:K}) associated with pure drawdown. We denote $z_0(t) =\phi\mathbb{E}[ I(t; \tilde{\mathbf{u}})]\!\,_tp_x$ and find the average cost function (\ref{eq:cost}) under (\ref{eq:gsa_eq}) as follows:
\begin{equation}\label{eq:pro}
\bar{C}_x^{(n)} = \int_0^{\bar{T}}z_0(t)\xi(t)\dd{t}.\end{equation}
As anticipated, we observe in (\ref{eq:pro}) that the provider will bear the risk arising from guarantee factor $\xi(t)$. Taking the perspective of the retiree, we denote $z_{\theta}(t) = (1+\theta)z_0(t)$ and discover in (\ref{eq:pro1}) a trade-off between the benefits of risk mitigation and the emergence of additional decumulation risks. We have
\begin{equation}\label{eq:pro1}
1 = X_{\angl{T}} +  \int_0^{\bar{T}} z(t)K(t;\mathbf{u})\dd{t} + \int_0^{\bar{T}} z_{\theta}(t)\left(1 - \dfrac{I(t; \mathbf{u})}{I(t; \tilde{\mathbf{u}})}\cdot \dfrac{\mathbbm{1}_{\{T > t\}}}{_tP_x^{(n)}}\cdot\dfrac{\xi(t)}{1 + \theta}\right) \dd{t}.
\end{equation}
\textit{Risk Mitigation}. In (\ref{eq:pro1}), we observe a form of \textit{longevity protection} to the retiree in the form of a stable payment stream at older ages, with the second integral occupying a smaller share of the retiree's initial wealth (or \$1) when $T$ is large. Indeed, \citet{Pi16} argues that the most important feature of life annuities to the retiree is protection against the risk of outliving their initial wealth $W_0$, reflecting a form of insurance against outliving expectations. The survival probability $\!\,_tP_x^{(n)}$ will fall over time as the membership ages, providing a rising stream of mortality credits to surviving participants in the pool \citep{AGA}. 

We observe the potential for \textit{portfolio variability reduction} through diversification across $\mathbf{u}$ and $\tilde{\mathbf{u}}$; and by recognising that mortality credits and market returns are effectively uncorrelated \citep{Bl18}, offering diversification benefits which cannot be replicated by traditional asset classes alone \citep{AlMa02}. While longevity protection will help satisfy a given liquidity risk tolerance $\psi_{\nu}$, this reduction in portfolio variability will be attractive to a retiree with high investment risk aversion $b$. \\

\noindent \textit{Risk Emergence}. In (\ref{eq:pro1}), we can set $T = T_x^*$ to observe the impact of liquidity risk in scenarios where $T_x^*$ takes a low value. The retiree will have effectively purchased a stream of future payments $\{d(t):t\in [0, \bar{T}]\}$ at prices $\{z_{\theta}(t): t\in [0, \bar{T}]\}$. However, there is a subset of payments $\{d(t): t\in (T_x^*,\bar{T}]\}$ that may neither be brought forward in event of early liquidity need nor bequeathed in event of death. Where $T_x^*$ is small, the retiree therefore pays for a larger subset of payments  $\{d(t): t\in (T_x^*,\bar{T}]\}$ that will never be received before first transition, that is the phase where survival status $\mathbbm{1}_{\{T_x^* > t \}}$ is non-zero.

Whether $T_x^*$ is considered late or early may be benchmarked against the inflated survival probability $(1 + \theta)\!\,_tP_x^{(n)}$ or $\left(1 + \theta\right)\!\,_tp_x$, depending on the form of guarantee factor $\xi(t)$. For instance, if the sequence of survival probabilities $\,(_tP_x^{(n)})_{t\in\mathcal{S}}$ decays faster over time than expected, then the retiree must not live as long to receive their money's worth from a GSA. That is, the second integral in \eqref{eq:pro1} will end up occupying a comparatively smaller share of the retiree's initial wealth $W_0$ (or \$1) if $\,(_tP_x^{(n)})_{t\in\mathcal{S}}$ decays faster for fixed $T_x^*$; or equivalently, $T_x^*$ is large for fixed $\,(_tP_x^{(n)})_{t\in\mathcal{S}}$. However, we see that the retiree effectively incurs a `longevity risk premium' \citep{Bl18}, with a higher loading factor $\theta$ increasing the amount of decay in $_tP_x^{(n)}$ that is required to achieve an effect of similar order.

\subsubsection{Possible Extensions}\label{sec:nat_hedge} 
\noindent We will now consider some possible extensions to the Annuity Family which might allow even finer tuning of the retiree's distribution of outcomes in line with their differing risk appetites. Though we do not implement these extensions in full, we seek to lay the groundwork for future work.  \\

\noindent \textit{Natural Hedging}. \citet{RaOg18} argue that an attractive annuity product must satisfy the risk appetite of the retiree whilst also promoting insurer sustainability. \citet{Ya65} recognises that buying a life annuity is broadly similar in concept to selling life insurance on your own lifetime $T_x$. Thus, the retiree may like to buy a \textit{death benefit} associated with their pooling policy in order to buy a \textit{natural hedge} against the early death risk associated with this pooling policy. 
\citet{ZhShXu22} find that the addition of a death benefit to an annuity to form a `bequest-enhanced annuity' should theoretially lower $\theta$ under Solvency II requirements. 

We can consider the purchase of a death benefit as a natural hedge in some death benefit ratio $\beta$ with payment rate $\phi$, achieving some lower loading $\theta_{\beta} \in [0, \theta]$. In \citet{De23}, natural hedging is implemented through integration of death benefit ratio $\beta$ into the payment vehicle, giving the \textit{bequest-enhanced} (BE) payment vehicle $g_{\beta}(t)$ as a function of $g(t) \equiv g_0(t)$. We have then
\begin{equation}{\label{eq:gsabe}}
g_{\beta}(t) = \begin{cases}g(t), &t < T_x; \\ g(t)\beta, & t = T_x. \end{cases} 
\end{equation}
While we do not implement natural hedging in this paper, the approach followed by \citet{De23} is provided in Appendix \ref{A_nathed}.\\

\noindent \textit{Deferred Annuity}. \citet{Ho08} propose the idea that the retiree may be able to benefit from the equity premium at younger ages and from exploiting higher mortality credits later in life. As such, \citet{Mi07} derive the optimal age $x+t^*$ to which a retiree at time $0$ should defer the annuitisation of their wealth. To perform this optimisation, we could write (\ref{eq:dd1}) as below:
\begin{equation} 
1 - P_{\theta} = X_{\angl{T}} + \int_0^{T} I(t; \mathbf{u})\big(c(t) - \phi \mathbbm{1}_{\{t > t^*\}} \big) \dd{t}.
\end{equation}
\noindent \textit{Combinations}. In practice, one can also mix payment streams to combine the risk mitigation benefits of some strategies with the flexibility of others \citep{AGA}. A sophisticated investor may want to optimise for some combination of $L$ payment streams $\{d_i(t):i=1,\dotsc,L\}$ with respective prices $\left\{P_i(t; \theta_i):i=1,\dotsc, L \right\}$ by writing (\ref{eq:dd1}) as (\ref{eq:pth}):
\begin{equation}\label{eq:pth}
1 - P_{\theta} = X_{\angl{T}} + \sum_{i=1}^{L} \int_0^{T} I(t; \mathbf{u})\big(c(t) - d_i(t)  \big) \dd{t},
\end{equation}
where 
\begin{equation}
P_{\theta} = (1 + \theta)\sum_{i=1}^{L} P_i(t; 0) \quad \ \ \text{for} \ \ \quad \theta =  \dfrac{\sum_{i=1}^{L} (1 + \theta_i) P_i(t; 0)}{\sum_{i=1}^{L} P_i(t; 0)}-1. \end{equation}
\textit{Age Pension}. The age pension may also be included in (\ref{eq:pth}) as the $j^{\text{th}}$ payment stream by $d_j(t) = m(t)$ with no premium upon retirement, that is $P_j(t; \theta_j) = 0$. For illustration, we consider the combination of the age pension and drawdown as a type of overall decumulation strategy in (\ref{eq:ap}):
\begin{equation}\label{eq:ap}
1 = X_{\angl{T}} + \int_0^{T} I(t; \mathbf{u})\big(c(t) - m(t)\big)\dd{t}.
\end{equation}
We can observe the similarity between (\ref{eq:ap}) and the form of $X_{\angl{T}}$ in (\ref{eq:pro0}) across the Annuity Family. For instance, an age pension that is fixed in real terms, so that $m(t)\equiv \bar{m}$, will function much like a zero-premium IIA. This kind of similarity may lead to the crowding out of private annuity markets, inhibiting their successful development \citep{Mi02}. That is, large government social security systems provide a supply of annuities that satisfies a large portion of demand, preventing pooling providers from attaining the economies of scale which make the pooling mechanism so effective.

\subsection{Set of Admissible Strategies}\label{sec:dec_v}
We find the overall decumulation strategy which maximises investment objective function in (\ref{eq:objj}) subject to liquidity risk constraint (\ref{eq:con}) from a set of admissible strategies. These strategies are to comprise some choice of investment account $A(\mathbf{u})$ and single payment stream $(d(t))_{t\in \mathcal{S}}$. We impose restrictions:
\begin{enumerate}
\item We permit the retiree to choose the trading strategy $\mathbf{u}$ of their investment account $A(\mathbf{u})$ under the restriction that $\mathbf{u}$ corresponds to holding some combination of the risk-free asset and the market portfolio. In particular, we define the \textit{market weight} $w \geq 0$ as the proportion of wealth remaining in investment account $A(\mathbf{u})$ which is allocated to the market portfolio.
\item We permit the retiree to choose a single payment stream $d(t) = \phi g(t)$ from the Annuity Family under the restriction that $\mathbb{E}[d(t)] \in[0, c(t)]$ for all $t < T_x$.  Further, we restrict our focus to $c(t)\equiv c$ where $c$ is given in real terms, which is consistent with the benchmarking approaches in Section \ref{sec:plan}. The drawdown strategy thus reduces into the (real) self-annuitisation case and we have the following equivalence across the Annuity Family: 
\begin{equation}\label{eq:phi_asss}
\mathbb{E}[d(t)] \leq c(t) \quad \iff \quad \phi \leq c.\end{equation} 
\item We permit the retiree to choose $(g(t))_{t\in \mathcal{S}}$ from Table \ref{table:taby}. For dimensionality reduction, we assume that trading strategy $\tilde{\mathbf{u}}$ is fully determined by the retiree's choice of $(g(t))_{t\in \mathcal{S}}$. We permit a portfolio return guarantee if and only if $\tilde{\mathbf{u}}$ is invested in the risk-free asset so that $\mathbb{E}[I(t; \tilde{\mathbf{u}})] \equiv I(t; \tilde{\mathbf{u}})$. Otherwise, we set $\tilde{\mathbf{u}} = \mathbf{u}$ even though pooling providers are likely to hold a more efficient portfolio than any individual investor.
\end{enumerate}

\noindent We thus implement the optimisation in Section \ref{sec:final_opt} over the admissible strategies summarised in Table \ref{table:adst}.
\begin{table}[hbt!]
$$\begin{array}{l||l|l}
  \text{\textbf{Strategy Component}} & \text{\textbf{Decision Variable}} & \text{\textbf{Permitted Set}} \\ \hline \hline
  \text{Investment Account} \ A(\mathbf{u}) & \text{Market Weight} \ w & w\geq 0  \\ \hline
  \text{Payment Stream} \ (d(t))_{t\in \mathcal{S}} & 
  \text{Payment Vehicle} \ (g(t))_{t\in \mathcal{S}}  & \text{Table \ref{table:taby}} \\
  &\text{Payment Rate} \ \phi & 0\leq \phi \leq c  \\
\end{array}
$$
\caption{Summary of Admissible Overall Strategies}
\label{table:adst}
\end{table}
\section{Approach to Implement the Optimisation Framework}\label{ch:6}
We seek to implement the optimisation framework of Section \ref{ch:3} over the admissible strategies found in Section \ref{sec:integ}. Our approach balances the trade-off between numerical precision of the optimum and computational speed through the following implementation approach:
\begin{enumerate}
    \item In Section \ref{sec:mod}, we prioritise numerical accuracy in the algorithm used to find the optimum;
    \item Across Section \ref{sec:theo_ass} and Section \ref{sec:TR}, we prioritise tractable assumptions to allow us to reduce run times.
\end{enumerate}

\subsection{Optimisation Algorithm}\label{sec:mod}
We propose an exhaustive search algorithm which is capable of producing high precision of results at the expense of longer run times. Numerical precision is important given the high dimensionality of our decision space in 
Section \ref{sec:dec_v}. First, we consider a retiree with consumption target $(c(t))_{t\in \mathcal{S}}$ and risk appetites that are parameterised by $b$ and $\psi_{\nu}$ for some choice(s) of $\nu$. We then analyse the sensitivity of the optimal strategy to the retiree's risk appetite and target SOL in order to determine the effect on decision-making and validate our framework.

\begin{enumerate}[label=(\arabic*)]
    \item \textit{Evaluating Components of Optimisation Problem}. 
    For each payment vehicle $(g(t))_{t\in \mathcal{S}}$ in Table \ref{table:taby}, we compute three grids of outputs $\mathbb{E}[X_x^{\mathcal{I}}]$, $\mathrm{Var}(X_x^{\mathcal{I}})$, $\mathbb{P}_{\mathcal{L}}(T_x^* \geq \tau_{\nu})$ from a single grid of inputs $(w, \phi)\in S$.
    \item \textit{Finding the Optimal Strategy}. For each payment vehicle $(g(t))_{t\in \mathcal{S}}$ in Table \ref{table:taby}, we 
    grid search to find the admissible combination $(w, \phi)\in S$ which achieves the maximum $Q_g^* = \max Q_x^{\mathcal{I}}$ subject to $\mathbb{P}_{\mathcal{L}}(T_x^* \geq \tau_{\nu}) \leq \psi_{\nu}$. We then determine the optimal overall strategy which achieves the global maximum $Q^* = \max Q_g^*$ across all payment vehicles ${(g(t))_{t\in\mathcal{S}}}$.
\item \textit{Analysing Sensitivities}. We find the optimal strategy across differing risk appetites $(b, \psi_{\nu})$ and consumption targets $(c(t) )_{t\in \mathcal{S}}$ in order to determine how the optimal strategy may vary from retiree to retiree. By analysing these sensitivities, we can also validate the robustness of our framework.
\end{enumerate}

In Algorithm \ref{alg:alg2}, we execute Step (2) by searching for the optimal strategy through the grids of outputs evaluated in Step (1). We can then apply Algorithm \ref{alg:alg2} for varying $(b, \psi_{\nu})$ and $(c(t))_{t\in \mathcal{S}}$ as per Step (3).
\begin{algorithm}
\caption{Finding the Optimal Strategy}\label{alg:alg2}
\KwResult{Optimal strategy $(g(t))_{t\in \mathcal{S}}$ and $(w,\phi)$ that achieves $Q^*$}{
  \For{$(g(t))_{t\in \mathcal{S}}$ in $\mathrm{Table \ \ref{table:taby}}$}{
  \For{$(w, \phi)$ in $S$}{
    $Q_x^{\mathcal{I}} = \mathbb{E}[X_x^{\mathcal{I}}] - b(\mathrm{Var}(X_x^{\mathcal{I}}) + \mathbb{E}[X_x^{\mathcal{I}}]^2 )$ \\
    \If{$\mathbb{P}_{\mathcal{L}}(T_x^* \geq \tau_{\nu}) \leq \psi_{\nu}$ $\mathbf{and}$ $Q_x^{\mathcal{I}} > Q^*_g$}{
    \textbf{set} $Q^*_{g} \coloneqq Q_x^{\mathcal{I}}$
}
}} \textbf{set} $Q^*\coloneqq\!\! \displaystyle\max_{(g(t))_{t\in\mathcal{S}}}\!\!Q_g^*$}
\end{algorithm}

\subsection{Theoretical Assumptions}\label{sec:theo_ass}
In practice, the optimisation algorithm described in Section \ref{sec:mod} can be computationally intensive. We therefore emphasise parsimony in our choice of theoretical assumptions in order to leverage powerful theoretical results in Section \ref{sec:TR}.

\subsubsection{Investment Assumption} Noting that the price level $\Pi(t)$ will drift upwards over time with some drift inflation rate $\pi$, we choose to project inflation dynamics using Standard Brownian Motion (SBM) $B_t^{\pi}$ on the natural filtration $\mathcal{F}_t$ as per, for instance, \citet{BrXi02}. We write
 \begin{equation}\label{eq:k}
\dd{\Pi(t)} = \pi \Pi(t)\dd{t} + \sigma_{\pi} \Pi(t)\dd{B_t^\pi}.\end{equation}
As suggested by \citet{DoKhLi23}, the inflation-indexed bond becomes the risk-free asset when a problem is expressed in real terms. Hence, $\mathbf{u}$ places weight $w$ on the market portfolio and weight $1 - w$ on inflation-indexed bonds which are assumed to deliver a risk-free rate of return $r$ in excess of inflation. We let $B_t^{\mu}$ be the SBM on $\mathcal{F}_t$ which drives the market portfolio and is assumed to be uncorrelated with $B_t^{\pi}$. Where $F(t; \mathbf{u}) = 1 / Y(t; \mathbf{u})$ is the nominal value of $\$1$ invested today at future time $t$, we have 
\begin{equation}\label{eq:1}
\dd{F(t; \mathbf{u})} = \mu_y(w)F(t; \mathbf{u})\dd{t} + \sigma_y(w)F(t; \mathbf{u})\dd{B_t^y},
\end{equation}
where
\begin{equation}\label{eq:nom}
\mu_{y}(w) = (1 - w)(r + \pi) + w \mu_M; 
\end{equation}
\begin{equation}\label{eq:exp}
\sigma_{y}(w)B_t^{y} = (1 - w)\sigma_{\pi}B_t^{\pi}+ w\sigma_M B_t^{\mu}.
\end{equation}
 
\subsubsection{Transition Assumption}\label{sec} 
As in \citet{Sh08}, we model the \textit{projected} death hazard rate over $\mathcal{S}$ as $\lambda_{x+t} = \lambda$ where
\begin{equation}
_tp_x = \dfrac{\mathrm{e}^{-\lambda t} - \mathrm{e}^{-\lambda \bar{T}}}{1 - \mathrm{e}^{-\lambda \bar{T}}}, \quad t\in [0, \bar{T}]. \end{equation}
We will let $\bar{T}\to \infty$ so that the distribution of $T_x$ approaches the exponential distribution $\mathrm{Exp}(\lambda)$ with rate parameter $\lambda$. We similarly assume $T_x^{\mathrm{ELN}} \overset{d}{=} \mathrm{Exp}(\lambda^{\text{ELN}})$, and make the simplifying assumption that $T_x^{\mathrm{ELN}}$ and $T_x$ are independent. 

Admittedly, the assumption of a constant force of mortality is a strong one, and it fails to account for the increasing convexity of survival probability $_tp_x$ over time. Nonetheless, this assumption will allow us to find a closed-form distribution for $X_x^{\mathcal{I}}$ in Section \ref{sec:TR} which allows greater computational efficiency in implementing the algorithm. While it is possible to relax this assumption, we would need to approximate the moments of $X_x^{\mathcal{I}}$ using Monte Carlo simulation.

\subsubsection{Population Assumption} We assume the population survival probability $_tp_x^{\mathrm{ref}}$ is differentiable with continuous hazard rate $\Lambda_{x+t}$. By extrapolating past mortality trends, \citet{LeCa92} project $\Lambda_{x+t}$ as stochastic into the future, being lognormal at each future time $t$. \citet{MiYo05} assume a diffusion process with lower bound $\underline{\lambda}$ reflecting some minimum mortality rate due to the hazard of accidents even after biological causes of death are removed.

We define population survival function $_tp_x^{\mathrm{ref}} = \mathrm{e}^{-\Lambda t}$ where $\Lambda - \underline{\lambda}$ is lognormally distributed with mean $\lambda - \underline{\lambda}$, and uncertainty parameter $\hat{\sigma}$. Though simplified, these dynamics showcase the `trend'-like nature of longevity risk discussed in Section \ref{sec:trans}. This is because the projected hazard rate $\lambda$ either overstates or understates $\Lambda$, with $_tp_x^{\mathrm{ref}}$ trending away from projected survival probability $_tp_x = \mathrm{e}^{-\lambda t}$ over time. 

When considering the participating policies GSA and LIA, we also let $_tP_{x}^{(n)} \to \, _tp_x^{\mathrm{ref}}$ by allowing the pool to become infinitely large. As suggested by \citet{BoFrRu23}, one could also model the mortality risk associated with finite pool size $n$ by drawing realisations for survivors over time from a suitable binomial distribution. Assuming lifetimes $\{T_x^{(i)}\}$ are independent, conditional on $_tp_x^{\mathrm{ref}}$, we can write $n\,_tP_x^{(n)} \overset{d}{=} \mathrm{Bi}(n,\, _tp_x^{\mathrm{ref}})$. By focusing only on longevity risk here, we potentially understate the impact of aggregate death risk on the retiree's decision-making, potentially making the GSA and LIA seem overly attractive.

\subsubsection{Loading Assumption}
In a simple setting without regulatory requirements, the pooling provider might target a form of Sharpe ratio $S$. \citet{MiPr06} develop a stylised framework for the pricing of aggregate death risk, accounting for the portfolio of a pooling provider with zero expenses. In \eqref{eq:sh0}, we simplify their expression for $S$ in terms of average cost function $\bar{C}_x^{(n)}$, and rearrange to find loading factor $\theta$.
\begin{equation}\label{eq:sh0}
S = \dfrac{nP_{\theta} - nP_0}{\sqrt{\mathrm{Var}\big(C_x^{(n)}\big)}} \quad \implies \quad \theta = \dfrac{S}{P_0}\sqrt{\mathrm{Var}\big(\bar{C}_x^{(n)}\big)}.
\end{equation}

\subsection{Theoretical Results}\label{sec:TR}
Since our optimisation algorithm is computationally intensive, we seek to construct tractable equations and theoretical results which reduce the need for simulation.

\subsubsection{Key Equations}\label{sec:KE} 
From (\ref{eq:k}) and (\ref{eq:1}), we first derive an expression for $I(t; \mathbf{u}) = Y(t; \mathbf{u})\Pi(t)$. Given $B_t$ is a SBM on the natural filtration $\mathcal{F}_t$ which depends on $B_t^{\pi}$ and $B_t^{\mu}$, we write 
\begin{equation}\label{eq:2}
I(t; \mathbf{u}) = \mathrm{e}^{-\left(\mu(w) - \frac{1}{2}\sigma(w)^2\right)t - \sigma(w) B_t},\end{equation}
where 
\begin{equation}\label{eq:mean_std}
\mu(w) = \mu_{y}(w) - \pi \qquad \text{and} \qquad 
\sigma(w)^2 = w^2\sigma_M^2 + w^2 \sigma_{\pi}^2. \end{equation}
We will consider values of $w$ such that the expected present value (EPV) of the lifetime annuity is well-defined under our model assumptions. That is, we require $\sigma(w)^2 < \lambda + \mu(w)$ so that we can write
\begin{equation}{\label{eq:ax}}
a_x(w) = \mathbb{E} \int_0^{T_{x}} I(t; \mathbf{u})\dd{t} = \dfrac{1}{\lambda + \mu(w) -\sigma(w)^2}.
\end{equation}
While (\ref{eq:ax}) can be surprising, it can be verified against \citet{MiRo05}. There exists market weight $w_0$ such that $a_x(w) \geq a_x(w_0)$ for all $w_0$. That is, the variability in the annuity present value increases in $w$ for $w > w_0$ without any improvement to the mean. When finding an optimal strategy, we therefore need only consider $w \leq w_0$ to avoid either purchasing or self-funding an annuity of given expected present value and excess variance. \\

\noindent The loading factor $\theta$ is determined from (\ref{eq:sh0}) based on $\bar{C}_x^{(n)}$. Given our population assumption, we can compute $\mathrm{Var}\big(\bar{C}_x^{(n)}\big)$ in (\ref{eq:sh0}) through  decomposition of variance:
\begin{equation}\label{eq:decom}
\mathrm{Var}\big(\bar{C}_x^{(n)}\big) =\mathbb{E}[\mathrm{Var}\big(\bar{C}_x^{(n)}\vert\Lambda\big)] + \mathrm{Var}(\mathbb{E}[\bar{C}_x^{(n)}\vert\Lambda]).
\end{equation}
From (\ref{eq:pro}), we have $\mathrm{Var}(\bar{C}_x^{(n)}) = 0$ for a GSA and LIA since all risk is transferred to the retiree. We can therefore take $\theta = 0$ in (\ref{eq:sh0}) for these payment streams.
\subsubsection{Expected Investment Utility}\label{sec:inv_ut}
We begin by applying information set $\mathcal{I}$ to (\ref{eq:pro0}). We have
\begin{equation}\label{eq:XI0}
1 -  (1 + \theta)P_0 = X_{x}^{\mathcal{I}}+ \int_0^{\infty} \mathrm{e}^{-\lambda t} I(t; \mathbf{u})\big(c(t) - \phi \hat{g}(t)\big)\dd{t}.\end{equation}
In information set $\mathcal{I}$, we consider population rates $_tp_x^{\mathrm{ref}}$ to be deterministic with hazard rate $\lambda$. Hence, we have $\hat{g}(t)= \mathbb{E}[I(t; \tilde{\mathbf{u}})] / I(t; \tilde{\mathbf{u}})$. Making use of the restrictions on $\tilde{\mathbf{u}}$ imposed in Section \ref{sec:dec_v} and $c(t) \equiv c$, we consider two cases for (\ref{eq:XI0}) across the Annuity Family. We remark that it may also be tractable to consider consumption targets of the form $c(t) = \mathrm{e}^{\varrho t}$ for $\varrho \in \mathbb{R}$.\\

\noindent \textit{Case 1} (IIA, LIA). We have $\tilde{\mathbf{u}}$ comprising only the risk-free asset. Hence, $\hat{g}(t) = 1$ giving 
\begin{equation}\label{eq:pow0}
\int_0^{\infty} \mathrm{e}^{-t\lambda} I(t; \mathbf{u})\big(c(t) - \phi \hat{g}(t)\big)\dd{t} =  \left(c - \phi\right)\int_0^{\infty} \mathrm{e}^{-\left(\lambda + \mu(w) - \frac{1}{2}\sigma(w)^2\right)t - \sigma(w) B_t} \dd{t}. \end{equation}
\textit{Case 2} (ULA, GSA). We have $\tilde{\mathbf{u}} \equiv \mathbf{u}$. Hence, $I(t;\mathbf{u})\hat{g}(t) = \mathbb{E}[I(t;\mathbf{u})]$ giving 
\begin{equation}\label{eq:pow1}
\int_0^{\infty} \mathrm{e}^{-t\lambda} I(t; \mathbf{u})\big(c(t) - \phi \hat{g}(t)\big)\dd{t} = c\int_0^{\infty} \mathrm{e}^{-\left(\lambda + \mu(w) - \frac{1}{2}\sigma(w)^2\right)t - \sigma(w) B_t}\dd{t} - \phi  a_x(w). \end{equation}
Through substituting (\ref{eq:pow1}) into (\ref{eq:XI0}), we can observe that the expected investment utility of a GSA is equal to that of pure drawdown for all market weights $w$, independent of the payment rate $\phi$. This is reasonable since both the GSA and pure drawdown are characterised by the same trading strategy $\mathbf{u}$ and zero loadings. Now, to compute the moments of $X_x^{\mathcal{I}}$, we utilise a result attributed to \citet{Du90} and applied by \citet{Mi97} in a drawdown context. They derive:
\begin{equation}\label{eq:cf0}
\left[\int_0^{\infty} \mathrm{e}^{-\left(v t + s B_t\right)}\dd{t}\right]^{-1} \ \overset{d}{=} \ \Gamma\left(\alpha=\dfrac{2v}{s^2},\, \eta=\dfrac{2}{s^2}\right).
\end{equation}
In (\ref{eq:cf0}), $\Gamma(\alpha,\eta)$ is a Gamma distribution with shape parameter $\alpha$, rate parameter $\eta$ and density function $f$ as per (\ref{eq:dens}):
\begin{equation}\label{eq:dens}
f(x) = \dfrac{\eta^{\alpha}x^{\alpha - 1}\mathrm{e}^{-\eta x}}{\Gamma(\alpha)}\mathbbm{1}_{\{x \geq 0\}}.
\end{equation}
It follows that the reciprocal of the common integral across (\ref{eq:pow0}) and (\ref{eq:pow1}) will follow a Gamma distribution with $v=\lambda+\mu(w)-\frac{1}{2}\sigma(w)^2$ and $s = \sigma(w)$. For $\alpha > N$, the $N^{\mathrm{th}}$ moment of the integral in (\ref{eq:pow0}) and (\ref{eq:pow1}) becomes $\eta^N \Gamma(\alpha- N) / \Gamma(\alpha)$, as applied by \citet{Mi98a} in their technique to price Asian options. It can be readily verified that the mean $\eta/(\alpha - 1)$ is precisely equal to annuity EPV $a_x(w)$. To use (\ref{eq:cf0}) to calculate $\mathbb{E}[(X_x^{\mathcal{I}})^2]$, we require $\alpha > 2$. Since $\alpha(w)$ is a decreasing function of $w$, we take the following upper bound:
\begin{equation}\label{eq:w_ceil}
w < \mathrm{min}(w_0, w_1) \quad \text{where} \quad \alpha(w_1) = 2.
\end{equation}
When (\ref{eq:w_ceil}) holds, the cost of funding the annuity will be mean-variance efficient, and we can find closed-form expressions for the moments of $X_x^{\mathcal{I}}$ across the Annuity Family. 

\subsubsection{Liquidity Shortfall Probability}\label{sec:lrrp}
We look to the literature to motivate a general expression for the liquidity shortfall probability $\mathbb{P}(T_x^*\geq \tau_{\nu})$. In the case of the pure drawdown strategy, \citet{Mi97} expressed nominal wealth $W_T$ at fixed time $T\equiv t^*$ as per (\ref{eq:nw}), under the assumption of nominal consumption at rate $c$. He writes
\begin{equation}\label{eq:nw}
W_T = W_0\, F(T; \mathbf{u})\left[1 - \int_0^T c Y(t; \mathbf{u})\dd{t} \right]. \end{equation}
\citet{Mi97} remarked from the monotonicity of the integral in (\ref{eq:nw}) that the retiree will experience the event of ruin before time $T$ if and only if the bracketed factor falls below zero. Under the $\mathcal{L}$ information set, it follows that:
\begin{equation}\label{eq:nw2}
\mathbb{E}[W_T \vert \mathcal{L}] = W_0 \mathrm{e}^{\mu_{y}(w) T}\left[1 - \int_0^T c\mathrm{e}^{-\mu_{y}(w) t}\dd{t} \right]. \end{equation}
We consider (\ref{eq:nw}) in our setting where $\phi\geq 0$ and $c$ is defined in real terms. We rewrite (\ref{eq:nw2}) in terms of real wealth $W_T^*$ for general $T\leq T_{x}$:
\begin{equation}\label{eq:rw}
\mathbb{E}[W_T^*\vert \mathcal{L}] = W_0 \mathrm{e}^{\mu(w)T}\left[1 - P_{\theta} - \int_0^T \Big(c - \phi\xi(t) \dfrac{_tp_x}{_tP_x^{(n)}}\Big)\mathrm{e}^{-\mu(w)t}\dd{t}\right]. \end{equation}
The bracketed term is precisely $X_x^{\mathcal{L}}$, with the ruin probability equal to the \rev{smallest non-negative} root $T = \tau_0$ of $X_x^{\mathcal{L}}=0$, or equivalently $\mathbb{E}[W_T^*\vert\mathcal{L}]=0$. In general, we consider the shortfall probability $\mathbb{P}_{\mathcal{L}}\!\left(T_x^*\geq \tau_{\nu}\right)$ where $T = \tau_{\nu}$ is the \rev{smallest} root of $X_x^{\mathcal{L}} = \nu$ as per (\ref{eq:nu}):
\begin{equation}\label{eq:nu} \nu\, =\, 1 - P_{\theta} - \int_0^T \Big(c - \phi\xi(t) \dfrac{_tp_x}{_tP_x^{(n)}}\Big)\mathrm{e}^{-\mu(w)t}\dd{t}.\end{equation}
Since $T_x^*\overset{d}{=}\mathrm{Exp}(\Lambda + \lambda^{\text{ELN}})$ in the $\mathcal{L}$ information set, we find the \textit{$\nu$\% shortfall probability}:
\begin{equation}\label{eq:condij}
\mathbb{P}_{\mathcal{L}}\!\left(T_x^*\geq \tau_{\nu}\right) = \mathbb{E}\big[\mathrm{e}^{- (\Lambda + \lambda^{\text{ELN}}) \tau_{\nu}}\big]. \end{equation}

\section{Implementation in the Australian Context}\label{sec:aus_con}
To illustrate our optimisation framework and implementation approach, we calibrate and apply them to the Australian context, and obtain interesting insights. These are discussed in details in Section \ref{ch:finds}.

\subsection{Calibrating Assumptions}\label{sec:aus_context}
We begin our implementation by calibrating the assumptions of Section \ref{sec:theo_ass} to the Australian context.
\subsubsection{Calibrating Hazard Rates to Australian Context}\label{sec:haz} We will consider a retiree of exact age $x = 67$, which is the age at which they first become eligible for the age pension in Australia. We calibrate $\lambda = 0.051$ to a 19.42 year life expectancy, which we derive for a male aged exactly 67. We draw upon the Australian Life Table 2019-21 mortality rates $q_{67+t}^{\text{ALT}}$ \citep{ABS}, paired with the most recent Australian Government Actuary (AGA) longevity improvement factors $f_{67+t}^{\text{AGA}}$ \citep{AGALT}, which are calibrated to the last 125 years of mortality improvement. Factors $f_{67+t}^{\text{AGA}}$ are given as a percentage annual rate of improvement in mortality rate $q_{67+t}^{\text{ALT}}$. In our choice of reference population, we implicitly assume there is no adverse selection of policies from the Annuity Family. \citet{AGALT} computes the one-year death probability in $t$ years time for an individual currently aged $67$:
\begin{equation}
q_{67+t} = q_{67+t}^{\mathrm{ALT}}\left(1 + \dfrac{f_{67+t}^{\mathrm{AGA}}}{100}\right)^t,\end{equation}
We find that the life expectancy of a 67 year old using the more recent, albeit more unreliable, AGA 25-year improvement factors is 20.30 compared to 19.42 derived from AGA 125-year improvement factors. We therefore choose to calibrate $\hat{\sigma} = 0.064$ by setting $\mathrm{Var}\left(1/\Lambda\right) = 1$, allowing for a one-year standard deviation of error in projections of life expectancy. We also set a minimum death hazard rate of $\underline{\lambda} = 0.010$. 

In our illustration, we set $T_{67}^{\mathrm{ELN}} \equiv T_{67}^{\mathrm{LTC}}$, that is we assume LTC needs are the only form of early liquidity needs prior to death. We use the HRS data tabulated by \citet{Br13} to calibrate $\lambda^{\text{LTC}} = 0.034$. We minimise the sum of squared differences between $\mathrm{e}^{-t\lambda^{\text{LTC}}}$ and the age-based proportions $_tp_{67}^{\text{HRS}}$ of initially healthy individuals requiring LTC over time, attaining a $R^2$ score of $83\%$.

\subsubsection{Calibrating Investment Returns}\label{sec:cal_inv} We assume a trend rate of inflation of $\pi = 2.5\%$ to reflect the midpoint of the target range of the Reserve Bank of Australia (RBA) of between 2.0\% and 3.0\% p.a. on average, over time \citep{rb22}. This rate is also broadly consistent with the 90-day average margin of 2.4\% as at 30 June 2023 in the yields between 10-year nominal bonds issued by the Australian Government and the corresponding inflation-indexed bonds. We calibrate annual inflation volatility $\sigma_{\pi}= 1.85\%$ based on the annualised mean square error in the quarterly growth rate of Consumer Price Index (CPI) around the $2.5\%$ target over the 30 years to June 2023. On behalf of the RBA, \citet{El22} states that the Australian long-term neutral cash rate is ‘at least’ 2.5\%. Given the current economic uncertainty, we allow for a reasonable risk-free rate of real return of $r=0.5\%$ through setting the yield on short-term liquid nominal bonds to $r+\pi = 3.0\%$. In an Australian context, \citet{Bi18} suggests that practitioners will often use an annual market risk premium of $6.5\%$ so we set $\mu_M = 9.5\%$. We calibrate annual market volatility $\sigma_M = 16\%$ according to the annualised standard deviation of daily log-returns derived for the S\&P/ASX 200 over the 10 years to 30 June 2023.

\subsubsection{Calibrating Loadings} Realistically, insurers will hold capital for both idiosyncratic longevity risk due to finite $n$ and systematic longevity risk \citep{Zh20}. As considered by \citet{MiPr06}, we find that setting $S = 20\%$ in (\ref{eq:sh0}) with $n=5000$ results in a reasonable loading factor $\theta = 18.3\%$ for the IIA. 

\citet{DuMaMi05} analysed the value of nominal annuities in Australia through the use of a Money’s Worth Ratio (MWR) which measures the cost effectiveness of a policy to the retiree by quantity $(1 + \theta)^{-1}$. Our simulated loading factor $\theta$ falls within the range of obtaining the MWR of 88\% derived by the authors for males aged 67, and the MWR of 66\% derived more recently by \citet{o15} for the IIA in an Australian context.

\subsubsection{Calibrating Consumption Targets} We calibrate values of constant consumption target $c(t) \equiv c$ based on the Annual Retirement Expenditure (ARE) figures given by the Association of Superannuation Funds of Australia (ASFA) for the June quarter 2023. Since ASFA grosses up weekly expenditure to arrive at ARE, and already accounts for price inflation, we can set $c = \text{ARE}/W_0$. Furthermore, ASFA sets a higher ARE for retirees aged 67-84 than retirees aged 85+, anticipating a reduction in daily spending needs at older ages. We therefore take a weighted average measure $\mathrm{ARE}^*$ of ARE as per (\ref{eq:adj2}) in order to avoid overstating spending needs whilst maintaining a constant consumption target.
\begin{equation}\label{eq:adj2}
\mathrm{ARE}^* = \dfrac{1}{3}\left(2\cdot \text{ARE}_{\text{67-84}} + \text{ARE}_{85+} \right).
\end{equation}
According to ASFA, the median superannuation account balance for Australians aged 60-64 was \$178,808 for males and $\$137,051$ for females \citep{Cl22}. However, a single homeowner seeking to retire at age 67 will require at least \$595,000 wealth at retirement to afford a Comfortable SOL, even still assuming some late-life dependence on the age pension \citep{ASFA}. Since 62\% of Australian retirees received the age pension as at June 2022 \citep{AIHW}, it is not appropriate to consider the median account balance in a setting without assuming the support of an age pension. We therefore set $W_0 = \$595,000$ in Table \ref{table:tabbb}.
 \begin{table}[hbt!]
 $$\begin{array}{r||c|c|c||c}
\text{SOL} & \text{ARE}_{\text{67-84}} & \text{ARE}_{\text{85+}} & \text{ARE*} & c  \\ \hline
\text{Modest} & \$31,867 & \$29,561 & \$31,098 & 5.2\%\\
\text{Comfortable} & \$50,207 &  \$46,788 & \$49,067  & 8.2\%
\end{array}$$
\caption{ASFA Consumption Rates}
\label{table:tabbb}
\end{table}

\noindent In addition, we are interested in observing the impact of the famous `4\% Rule' \citep{Be94} which sets annual consumption to 4\% of \textit{initial} wealth, adjusted for inflation over time. Based on an empirical study, the Bengen 4\% Rule was proposed to maintain the real-value of first-year consumption over a 30 year planning horizon in 90\% of scenarios \citep{BeDo18}. Overall, we consider consumption targets $c\in \left\{4.0\%, 5.2\%, 8.2\%\right\}$.

 \subsection{Notes on Algorithm Implementation}\label{sec:cass}
In Section \ref{sec:mod}, we introduced an optimisation algorithm to find the optimal strategy across retirees. We now discuss some practical notes regarding implementation:

\begin{enumerate}
    \item Since $_tP_{67}^{(5000)}$ is non-decreasing, we observe in (\ref{eq:nu}) that $X_{67}^{\mathcal{L}}$ is not necessarily monotonic in $T$ for the GSA and LIA, sometimes having multiple non-negative roots. We therefore partition $[0, 80]$ into a series of 10 intervals and conduct a search over each successive interval until the first root $\tau_{\nu}$ is found.
    \item Given some chosen resolution of grid for $(b, \psi_{\nu})$, we seem to require some higher resolution of grid for $(w, \phi)$ to reduce noise around decisions between payment vehicles $(g(t))_{t\in \mathcal{S}}$ for local values of $(b, \psi_{\nu})$. We find that a $40\times 40$ grid resolution for $(b, \psi_{\nu})$ under our calibrated assumptions will require at least a $750\times 750$ grid resolution for $(w, \phi)$. As $\sigma$ rises, we also notice that the required resolution falls as the trade-off between mean and variance of bequest becomes more distinct.
    \item The expressions for $\mathrm{Var}(\bar{C}^{(5000)}_{67})$ 
in (\ref{eq:decom}) and $\mathbb{P}(T_{67}^* \geq \tau_{\mathcal{\nu}})$ in (\ref{eq:condij}) arise from conditional probability, requiring iteration over the values of $\Lambda$. Both these quantities are also functions of $w$, so we must perform these calculations repetitively as we generate our grids of values. We take a relatively low sample size of $200$ for $\Lambda$ to offset some computational expense.
\item The expected funding cost  of lifetime annuity $a_{67}(w)$, computed from (\ref{eq:ax}) as a present value, is a convex function of $w$ with minimum at $w_0 = 1.25$. As discussed in \ref{sec:inv_ut}, $\mathrm{Var}(X_{67}^{\mathcal{I}})$ will be the variance of an inverse gamma distribution for all weights less than $w_1 = 2.30$. Following from (\ref{eq:w_ceil}), it becomes unnecessary to consider a higher market weight than $w_0 = 1.25$. It is also unlikely that retirees can achieve market weights exceeding 1.
\end{enumerate}

\section{Findings}\label{ch:finds}
We now present the results of implementing our optimisation framework in the Australian context and discuss the key implications for the Australian annuity market and the broader validity of our framework. 

\subsection{Results}\label{sec:res}
Overall, we find that the retiree's optimal choice of overall decumulation strategy will depend on the retiree's investment risk aversion (IRA) and liquidity risk tolerance (LRT) when selecting from admissible payment vehicles $(g(t))_{t\in \mathcal{S}}$ and decision variables $(w, \phi)$. We compare these decisions for different choices of consumption target $c\in \left\{4.0\%, 5.2\%, 8.2\%\right\}$ and shortfall threshold $\nu\in \{0\%, 20\%\}$. Where $\nu = 0\%$, the retiree is said only to have limited appetite for the probability of ruin in the $\mathcal{L}$ information set.

\begin{figure}[hbt!]    
\begin{center}
        \caption{Decision Boundaries across Annuity Family}\label{fig:solver0}
\footnotesize{The optimal payment vehicle $(g(t))_{t\in \mathcal{S}}$ is determined for varying IRA and LRT. Whitespace indicates that there are no admissible strategies within the retiree's appetite for liquidity risk. The highest level of risk aversion is in the top left corner.}\end{center}

    \begin{subfigure}[t]{\linewidth}
        \centering
        \caption{0\% Shortfall Threshold}\includegraphics[width=0.85\textwidth]{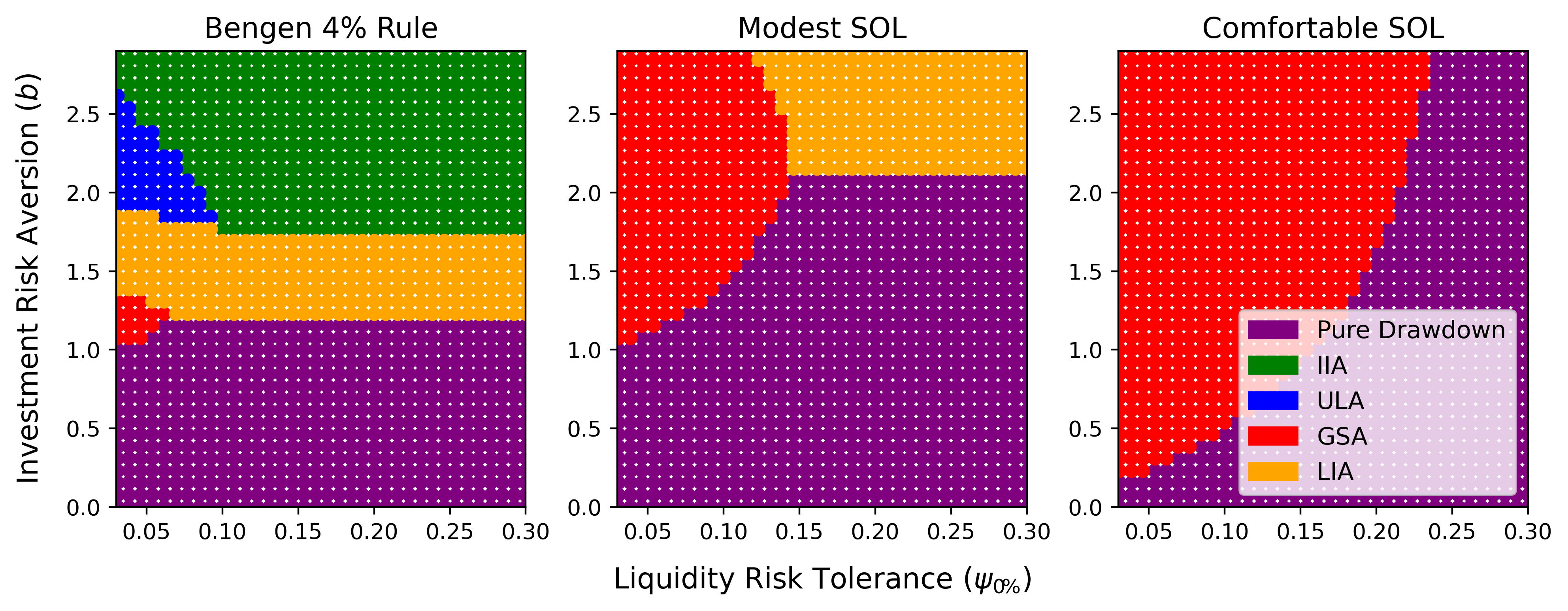}
    \end{subfigure}%
    
    \begin{subfigure}[t]{\linewidth}
        \centering
        \caption{20\% Shortfall Threshold}\includegraphics[width=0.85\textwidth]{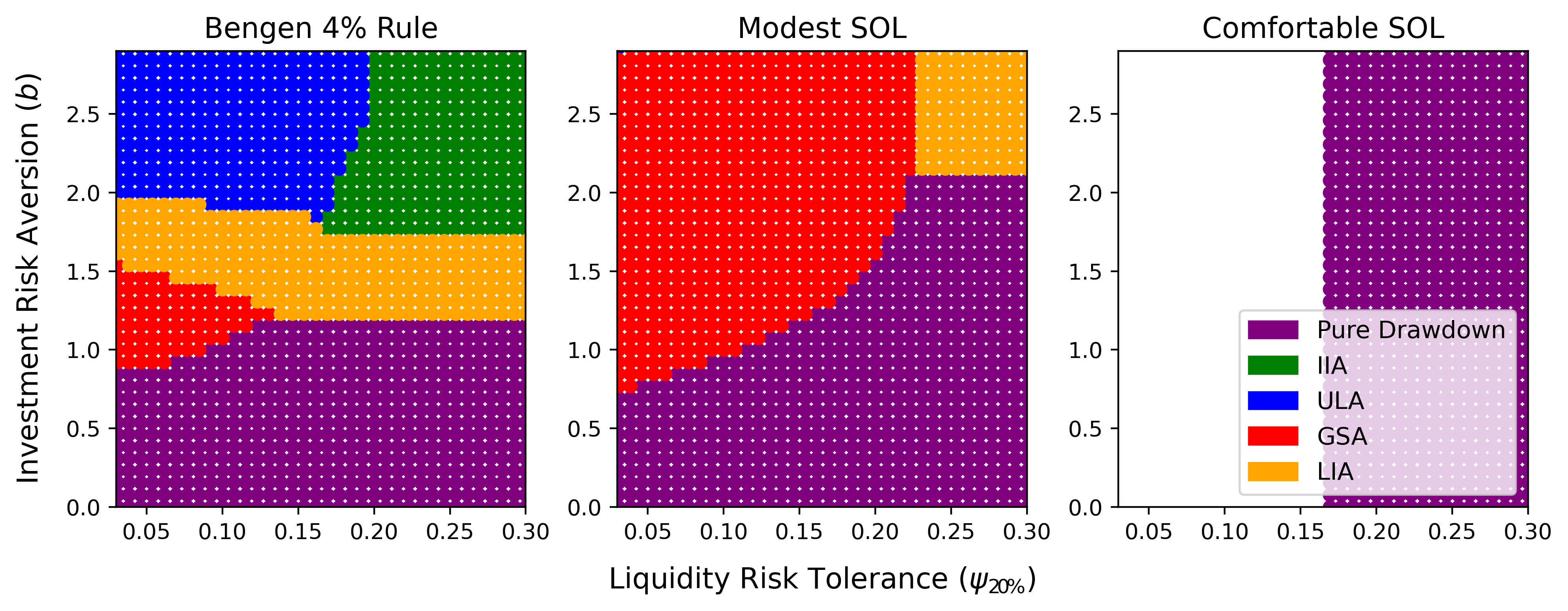}
    \end{subfigure}
\end{figure}

In Figure \ref{fig:solver0}, we find evidence of optimal strategic choice across the Annuity Family by determining the optimal choice of pooling policy (if any) associated with each retiree's IRA and LTR. It is optimal for retirees to purchase a payment stream from the Annuity Family for the purpose of portfolio variability reduction in the case that they have little appetite for investment risk; or longevity protection in the case they have little appetite for liquidity risk. In particular, retirees with little appetite for both decumulation risks will correspond to model points in the top left-hand corner of the plot associated with their chosen consumption target $c$ and shortfall threshold $\nu$.  

For lower values of IRA, the preference for the group self-annuitisation (GSA) over both the unit-linked annuity (ULA) and inflation-indexed annuity (IIA) may be attributed to the higher expected bequest due to the impact of zero loadings. For high IRA, a rising preference for the longevity-indexed annuity (LIA) and the IIA can be attributed to the valuable benefits of portfolio variability reduction associated with a lower allocations of wealth to the market portfolio.

\begin{figure}[hbt!]
\begin{center}
\caption{Liquidity Shortfall Probability for GSA}\label{fig:RP0}
\footnotesize{Liquidity shortfall probabilities $\mathbb{P}_{\rev{\mathcal{L}}}(T_{67}^*\geq \tau_{\nu})$ are plotted for varying target SOL, payment rate $\phi$ and market weight $w$. A shortfall threshold of $\nu = 20\%$ is taken. To illustrate dynamics, the axes are rescaled across target SOL.} \end{center}
\centerline{\includegraphics[width=0.7\linewidth]{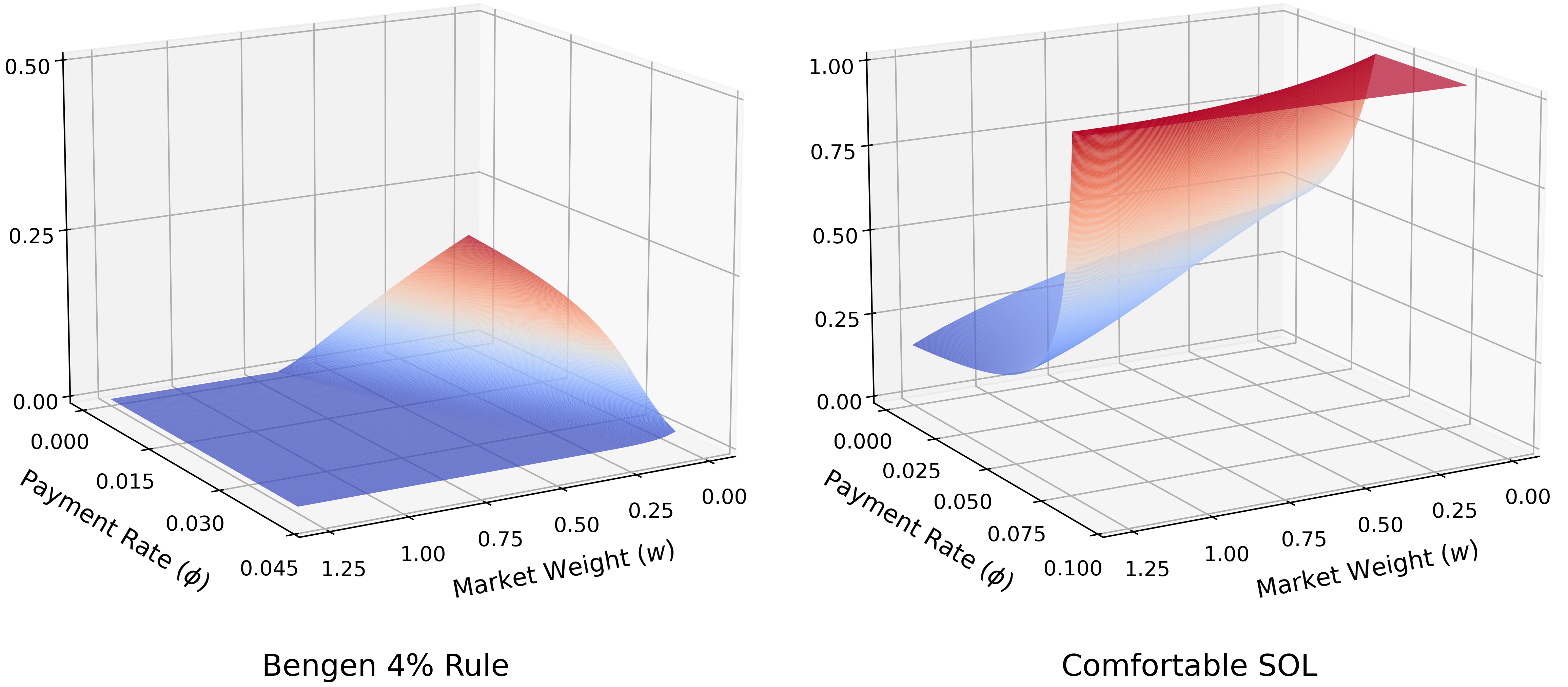}}
\end{figure}

For the Bengen 4\% Rule, we observe in Figure \ref{fig:RP0} that the GSA is able to achieve a lower shortfall probability than pure drawdown (i.e. $\phi = 0$) at larger choices of market weight $w$. For less ambitious consumption targets, the retiree will retain enough wealth to delay shortfall long enough to receive the longevity protection discussed in Section \ref{sec:prot}. However, the GSA becomes less viable in achieving a given level of LRT, as the shortfall probability begins to rise in the payment rate $\phi$. For a retiree seeking to consume at a comparably high rate relative to their initial wealth $W_0$, we deduce that the premium $P_{\theta}$ of the GSA can drain their savings, leaving them exposed to higher risk of shortfall than for pure drawdown.

\begin{figure}[hbt!]
\begin{center}
   \caption{Parameters of the Optimal Strategy} \label{fig:rainbow}
{\footnotesize{The optimal payment rate $\phi$ and market weight $w$ are plotted for varying IRA. A LRT of $\psi_{20\%} = 0.2$ is taken.}}\end{center}
    {\centering
    \begin{subfigure}[hbt!]{\linewidth}
        \centering
        \caption{Bengen 4\% Rule}
        \includegraphics[width=0.70\textwidth]{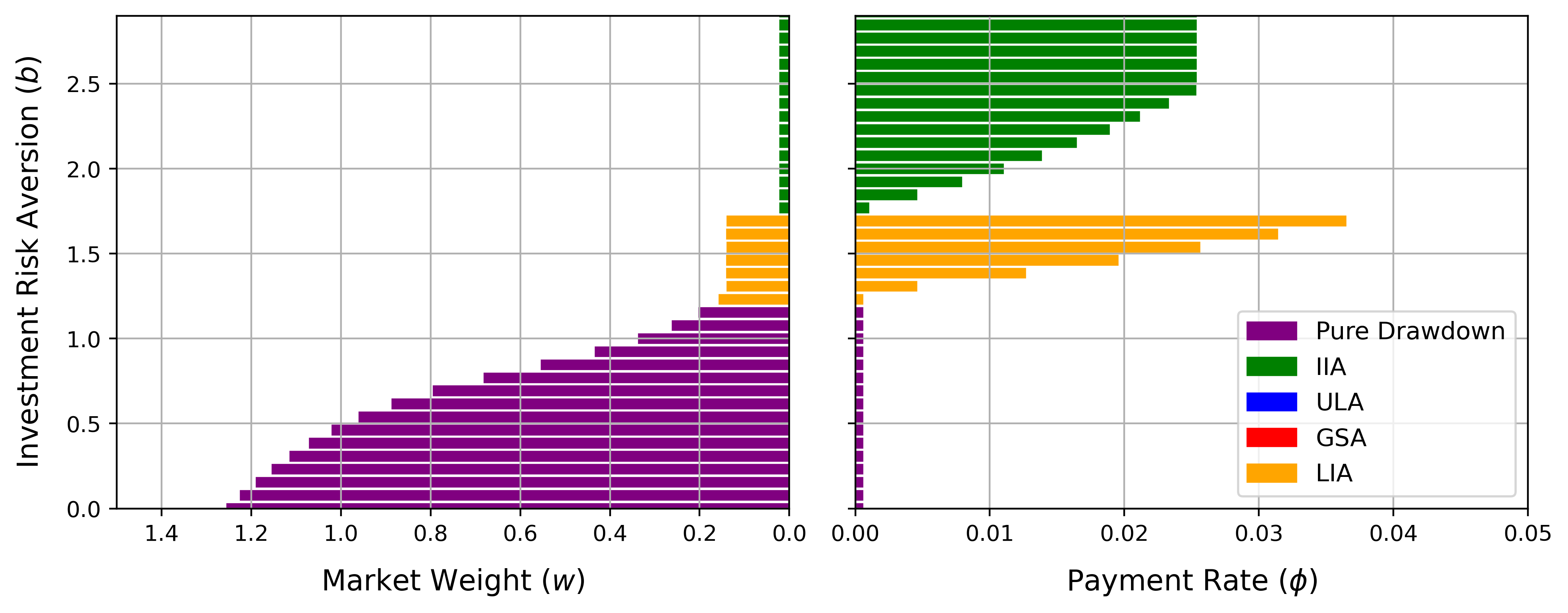}
    \end{subfigure}%
    
    \begin{subfigure}[hbt!]{\linewidth}
        \centering
        \caption{Modest SOL}
        \includegraphics[width=0.70\textwidth]{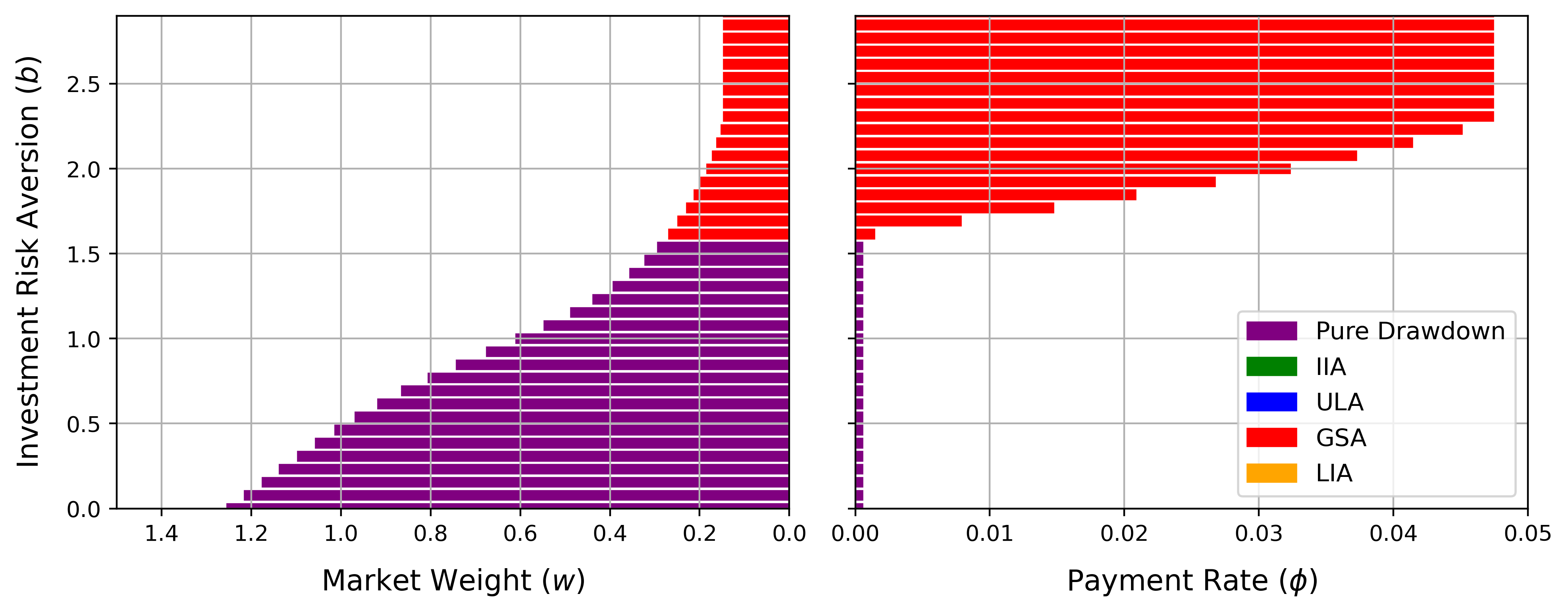}
    \end{subfigure}}
\end{figure}

In \citet{De23}, the retiree was given the ability to complement their choice of pooling policy from the Annuity Family with a death benefit. Although there was no qualitative change to Figure \ref{fig:solver0}, a retiree under the Bengen 4\% Rule was shown to purchase increasing amounts of death benefit as the IRA rose. This adoption suggests that a retiree with a less ambitious consumption target may be willing to trade off some surplus assets to reduce their variance of bequest. However, since the death benefit will effectively increase the premium $P_{\theta}$, the retiree largely preferred to self-fund their bequest as their consumption target became more ambitious. 
\newpage Figure \ref{fig:solver0} only provides insight into the optimal choice of payment vehicle $(g(t))_{t\in \mathcal{S}}$. In Figure \ref{fig:rainbow}, we complete the overall decumulation strategy by finding the optimal choice of $(w, \phi)$ for a retiree with given IRA and LRT. As the IRA rises, we learn that the distinct clusters in Figure \ref{fig:solver0} under the Bengen 4\% Rule and Modest SOL largely reflect a reallocation of wealth away from risky market-based strategies with high market weight $w$ towards less variable forms of longevity protection with increasing payment rate $\phi$.

\newpage To provide a solution to the decumulation problem, we transform the parameters of the optimal strategy in Figure \ref{fig:rainbow} into optimal allocations of initial wealth $W_0$. That is, we have the following \textit{transformation}:
\begin{enumerate}
\item Take $P_{\theta}$ as the initial allocation of wealth to a choice of pooling policy from the Annuity Family; 
\item Take $w(1 - P_{\theta})$ as the initial allocation of wealth to the market portfolio in the case of IIA and ILA; and take $w$ otherwise. Wealth not allocated to the market portfolio is allocated to the risk-free asset.
\end{enumerate} 

\begin{figure}[hbt!]

\begin{center}
{
   \caption{Parameters of Optimal Strategy}\label{fig:rainbow1}
   \footnotesize{The optimal allocations of initial wealth $W_0$ are plotted for varying IRA. A LRT of $\psi_{20\%} = 0.2$ is taken.}}

\end{center}
    {\centering
    \begin{subfigure}[hbt!]{\linewidth}
        \centering
        \caption{Bengen 4\% Rule}
        \includegraphics[width=0.7\textwidth]{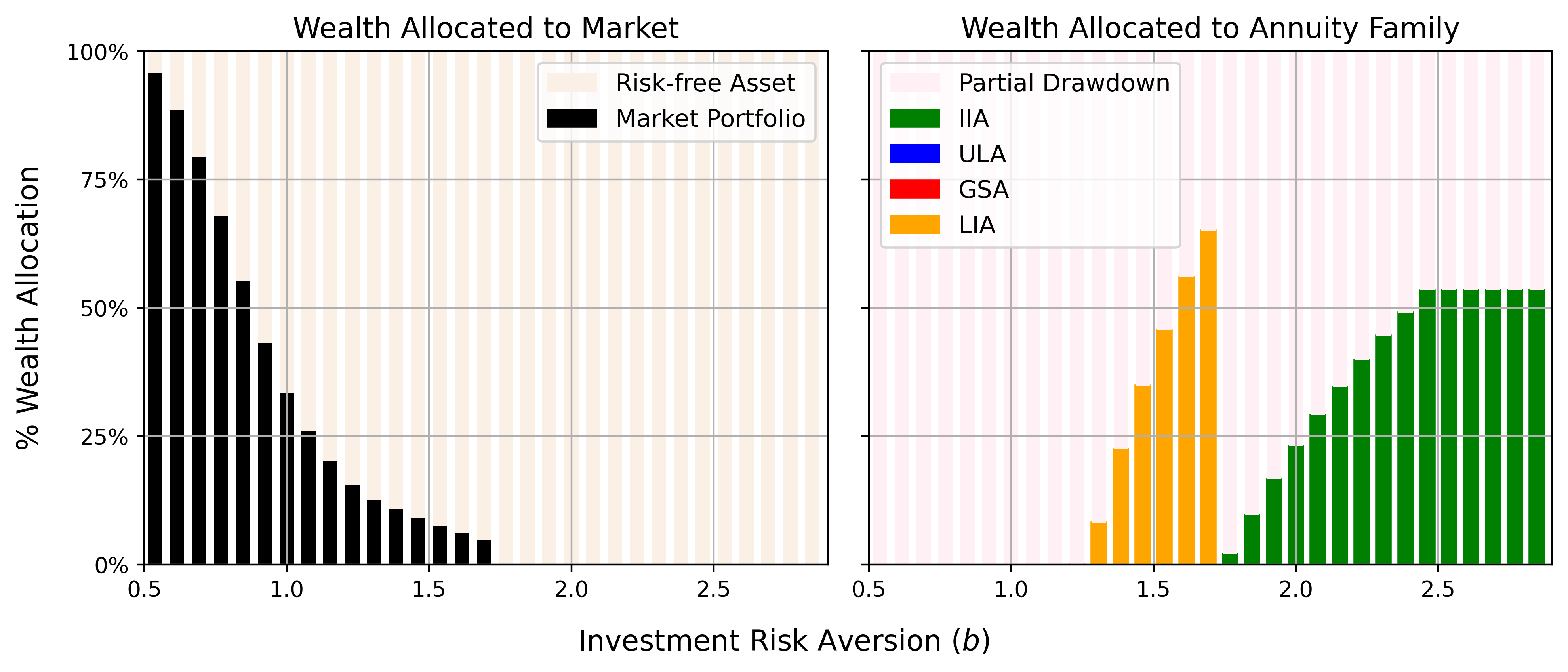}
    \end{subfigure}%
    
    \begin{subfigure}[hbt!]{\linewidth}
        \centering
        \caption{Modest SOL}
        \includegraphics[width=0.7\textwidth]{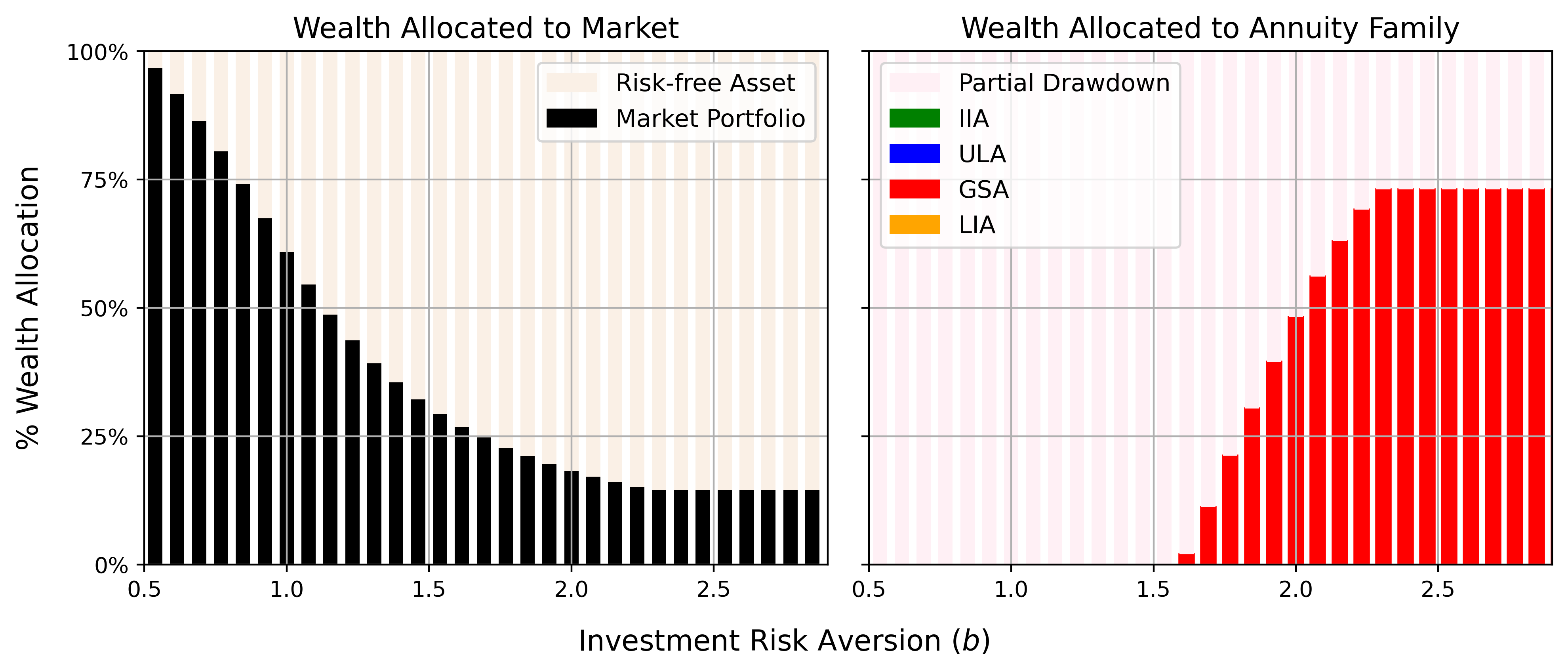}
    \end{subfigure}}
\end{figure}

In Figure \ref{fig:rainbow1}, we only consider wealth allocations of at most 100\% by restricting $b\geq 0.5$ so that $w\leq 1$ in Figure \ref{fig:rainbow}. That is, we do not consider optimal strategies at relatively low levels of IRA which involve short-selling the risk-free asset.

For the Bengen 4\% Rule, the jagged progression of market weight $w$ in Figure \ref{fig:rainbow} is smoothed out by the transformation in Figure \ref{fig:rainbow1}. However, we still observe a sudden jump in the wealth allocated across the Annuity Family. The retiree allocates decreasing amounts of wealth to the market in order to satisfy an increasing IRA. At lower expected portfolio returns, the retiree purchases increasing amounts of longevity protection via the LIA in order to remain within a given LRT. As the marginal benefits of longevity protection diminish, the retiree effectively opts for a cash-based drawdown strategy where $(w, \phi) \simeq (0,0)$ before finally accepting the non-zero loadings of the IIA as a way of decreasing their absolute variance of bequest. By permitting the retiree to purchase combinations from the Annuity Family as per Section \ref{sec:nat_hedge}, we might expect a jump of this nature to disappear as the portfolio is more seamlessly reallocated from product to product.

\newpage \subsection{Discussion}
\subsubsection{Australian Context}
  \rev{We find evidence to suggest it is optimal for many Australian retirees with \$595,000 of initial wealth to choose a pure drawdown strategy. Said differently, the optimal choice of decumulation strategy is pure drawdown over a large range of potentially differing appetites for liquidity and investment risks. Furthermore, t}he tendency to choose the pure drawdown strategy will grow as the retiree's target standard of living becomes more ambitious. However, an Australian retiree who is particularly averse to liquidity risk and/or investment risk may choose a payment stream from the Annuity Family. We observe that there is an allocation of wealth away from risky market-based strategies towards forms of longevity \rev{protection as investment risk appetite rises for some given liquidity risk appetite.} 

\rev{As discussed in Section \ref{sec:aus_context}, the median account balance in Australia upon retirement is actually significantly lower than the \$595,000 assumed.} Since we considered consumption target $c(t) \equiv c$ as a proportion of initial wealth, most Australians would therefore be forced to choose relatively ambitious consumption targets in pursuit of even a Modest SOL. Hence, we deduce that it is optimal for the majority of Australians to choose pure drawdown as their overall decumulation strategy, with annuities proving only optimal for the rich.

\citet{Ya65} put forward the seminal proposition: when individuals face only longevity risks, their savings should be completely annuitised in the absence of a bequest motive. In reality, there is a `puzzling' phenomenon of shallow annuity markets outside this stylised setting \citep{Mo86}, with a relatively small number of individuals voluntarily purchasing annuity products when they enter retirement age \citep{Pe17}, suggesting that annuities are actually far from optimal. Our deductions are therefore consistent with the phenomenon of a shallow annuity market, which is especially prominent in the Australian context \citep{o15}. We find that the non-purchase of annuities can be rationalised in line with the retiree's varying risk aversions, and according to our modelling.

\subsubsection{Broader Context}
Through the illustration of the Australian context, we find significant evidence suggesting that the optimal choice of overall strategy will change depending on the retiree's potentially varying aversions towards investment risk and liquidity risk. Moreover, the smoothness of the plots in Section \ref{sec:res} testifies to the power of the theoretical results in Section \ref{sec:TR} which included a closed-form distribution for $X_x^{\mathcal{I}}$. These theoretical results allowed us to not just reduce some computational expense, but also to improve the numerical precision of the optimum by avoiding the noise of simulation. Overall, our results indicate that there is value in broader application and extension of our optimisation framework.

\section{Conclusion}
We developed a framework to find the decumulation strategy which optimises the distribution of outcomes against a retirement plan according to the retiree's differing appetites for liquidity and investment risks. We found the optimal strategy amongst the Annuity Family, allowing for combination with drawdown. We proposed an approach to efficiently implement our optimisation framework, which facilitated for comparison of the optimal strategy from retiree to retiree.
Under the illustrative example of the Australian context, our methodology suggests that it might well be optimal for an Australian retiree to choose pure drawdown without purchasing any annuity. These findings are consistent with the shallow annuity market in Australia, lending support to our framework and indicating that there is value in broader application and extension of our optimisation framework. We also concluded that the optimal strategy for a retiree is likely to be a non-trivial function of the retiree’s differing risk appetites.

However, we believe that it is important to allow for the age pension in our set of admissible strategies, laying the groundwork for implementation in Section \ref{sec:nat_hedge}. Though we can also consider combination of strategies from the Annuity Family, there is a tension between complexity of strategy and the ability for the retiree to interpret their distribution of outcomes. In particular, we believe that only very sophisticated investors would consider integrating more than one payment stream from the Annuity Family into their overall decumulation strategy. Rather than expand the set of admissible strategies, further work to improve the efficiency of the optimisation algorithm would permit the valuable inclusion of more complex model assumptions such as age-dependent hazard rates. Through allowing $T_{x}$ to vary in distribution by age $x$, one may even be able to observe the effect of ageing on the optimal decumulation strategy as decisions are made dynamically over time.

\section*{Dedication}

This paper is dedicated to the eminent Professor Ermanno Pitacco, whose many seminal contributions in the areas of retirement, longevity, and mortality have profoundly influenced the field. The insights and knowledge shared by Professor Pitacco, notably through his frequent visits to Australia, have inspired the theoretical and illustrative contributions of this paper.

\section*{Data and Code}
The Python script that was used to obtain the results in Section \ref{sec:res} is available on \url{https://github.com/agi-lab/decumulation-risk}. The run-time is around 80 hours using a 2.6 GHz 6-Core Intel i7 processor.

\section*{Funding \rev{ and Acknowledgments}}
Avanzi acknowledges financial support under Australian Research Council's Discovery Project DP200101859 funding scheme. The views expressed herein are those of the authors and are not necessarily those of the supporting organisation. 

\rev{The authors are grateful to two anonymous referees, whose comments improved the presentation of the paper.}

\section*{Conflicts of interest declarations}
In 2022 and 2023, De Felice was employed by KPMG, a consultant active in the area of Australian superannuation. \rev{Although this is not the case at time of submitting this paper, both authors may look to work or collaborate within the superannuation industry in the future.}

\rev{The authors have no further potential conflicts of interest to declare.}

\section*{References}
\bibliographystyle{elsarticle-harv}
\bibliography{libraries}

\begin{thebibliography}{73}
\expandafter\ifx\csname natexlab\endcsname\relax\def\natexlab#1{#1}\fi
\providecommand{\url}[1]{\texttt{#1}}
\providecommand{\href}[2]{#2}
\providecommand{\path}[1]{#1}
\providecommand{\DOIprefix}{doi:}
\providecommand{\ArXivprefix}{arXiv:}
\providecommand{\URLprefix}{URL: }
\providecommand{\Pubmedprefix}{pmid:}
\providecommand{\doi}[1]{\href{http://dx.doi.org/#1}{\path{#1}}}
\providecommand{\Pubmed}[1]{\href{pmid:#1}{\path{#1}}}
\providecommand{\bibinfo}[2]{#2}
\ifx\xfnm\relax \def\xfnm[#1]{\unskip,\space#1}\fi
\bibitem[{ABS(2022)}]{ABS}
\bibinfo{author}{ABS}, \bibinfo{year}{2022}.
\newblock \bibinfo{title}{Life {T}ables}.
\newblock
  \bibinfo{howpublished}{\url{https://www.abs.gov.au/statistics/people/population/life-tables/latest-release}}.
\bibitem[{AIHW(2023)}]{AIHW}
\bibinfo{author}{AIHW}, \bibinfo{year}{2023}.
\newblock \bibinfo{title}{Income {S}upport for {O}lder {A}ustralians}.
\newblock
  \bibinfo{howpublished}{\url{https://www.aihw.gov.au/reports/australias-welfare/income-support-older-australians}}.
\bibitem[{Albrecht and Maurer(2002)}]{AlMa02}
\bibinfo{author}{Albrecht, P.}, \bibinfo{author}{Maurer, R.},
  \bibinfo{year}{2002}.
\newblock \bibinfo{title}{{S}elf-annuitization, {C}onsumption {S}hortfall in
  {R}etirement and {A}sset {A}llocation: {T}he {A}nnuity {B}enchmark}.
\newblock \bibinfo{journal}{Journal of Pension Economics \& Finance}
  \bibinfo{volume}{1}, \bibinfo{pages}{269--288}.
\bibitem[{ASFA(2023)}]{ASFA}
\bibinfo{author}{ASFA}, \bibinfo{year}{2023}.
\newblock \bibinfo{title}{Retirement {S}tandard}.
\newblock
  \bibinfo{howpublished}{\url{https://www.superannuation.asn.au/resources/retirement-standard}}.
\bibitem[{B\"{a}r and Gatzert(2022)}]{BaGa22}
\bibinfo{author}{B\"{a}r, M.}, \bibinfo{author}{Gatzert, N.},
  \bibinfo{year}{2022}.
\newblock \bibinfo{title}{{P}roducts and {S}trategies for the {D}ecumulation of
  {W}ealth during {R}etirement: {I}nsights from the {L}iterature}.
\newblock \bibinfo{journal}{North American Actuarial Journal}
  \bibinfo{volume}{27}, \bibinfo{pages}{322--340}.
\bibitem[{Bayraktar and Young(2008)}]{BaYo12}
\bibinfo{author}{Bayraktar, E.}, \bibinfo{author}{Young, V.},
  \bibinfo{year}{2008}.
\newblock \bibinfo{title}{{M}aximizing {U}tility of {C}onsumption {S}ubject to
  a {C}onstraint on the {P}robability of {L}ifetime {R}uin}.
\newblock \bibinfo{journal}{Finance Research Letters} \bibinfo{volume}{5},
  \bibinfo{pages}{204--212}.
\bibitem[{Bengen(1994)}]{Be94}
\bibinfo{author}{Bengen, W.}, \bibinfo{year}{1994}.
\newblock \bibinfo{title}{{D}etermining {W}ithdrawal {R}ates using {H}istorical
  {D}ata}.
\newblock \bibinfo{journal}{Journal of Financial Planning} \bibinfo{volume}{7},
  \bibinfo{pages}{171--180}.
\bibitem[{Bernhardt and Donnelly(2018)}]{BeDo18}
\bibinfo{author}{Bernhardt, T.}, \bibinfo{author}{Donnelly, C.},
  \bibinfo{year}{2018}.
\newblock \bibinfo{title}{Pension Decumulation Strategies: A State-of-the-art
  Report}.
\newblock \bibinfo{publisher}{Risk Insight Lab, Heriot-Watt University}.
\bibitem[{Bishop et~al.(2018)Bishop, Carlton and Pan}]{Bi18}
\bibinfo{author}{Bishop, S.}, \bibinfo{author}{Carlton, T.},
  \bibinfo{author}{Pan, T.}, \bibinfo{year}{2018}.
\newblock \bibinfo{title}{{M}arket {R}isk {P}remium: {A}ustralian {E}vidence}.
\newblock Business Valuation Reseach Paper, \bibinfo{publisher}{Chartered
  Accountants Australia \& New Zealand}.
\bibitem[{Blake(2018)}]{Bl18}
\bibinfo{author}{Blake, D.}, \bibinfo{year}{2018}.
\newblock \bibinfo{title}{{L}ongevity: a {N}ew {A}sset {C}lass}.
\newblock \bibinfo{journal}{Journal of Asset Management} \bibinfo{volume}{19},
  \bibinfo{pages}{278--300}.
\bibitem[{Boido and Fasano(2023)}]{BoFa23}
\bibinfo{author}{Boido, C.}, \bibinfo{author}{Fasano, A.},
  \bibinfo{year}{2023}.
\newblock \bibinfo{title}{{M}ean-variance {I}nvesting with {F}actor {T}ilting}.
\newblock \bibinfo{journal}{Risk Management} \bibinfo{volume}{25},
  \bibinfo{pages}{1--24}.
\bibitem[{Bommier and Villeneuve(2012)}]{BoVi12}
\bibinfo{author}{Bommier, A.}, \bibinfo{author}{Villeneuve, B.},
  \bibinfo{year}{2012}.
\newblock \bibinfo{title}{Risk {A}version and the {V}alue of {R}isk to {L}ife}.
\newblock \bibinfo{journal}{Journal of Risk and Insurance}
  \bibinfo{volume}{79}, \bibinfo{pages}{77--104}.
\bibitem[{B\"{o}rger et~al.(2023)B\"{o}rger, Freimann and Ruß}]{BoFrRu23}
\bibinfo{author}{B\"{o}rger, M.}, \bibinfo{author}{Freimann, A.},
  \bibinfo{author}{Ruß, J.}, \bibinfo{year}{2023}.
\newblock \bibinfo{title}{{O}n the {E}conomics of the {L}ongevity {R}isk
  {T}ransfer {M}arket}.
\newblock \bibinfo{journal}{Journal of Risk and Insurance}
  \bibinfo{volume}{90}, \bibinfo{pages}{597--632}.
\bibitem[{Brennan and Xia(2002)}]{BrXi02}
\bibinfo{author}{Brennan, M.}, \bibinfo{author}{Xia, Y.}, \bibinfo{year}{2002}.
\newblock \bibinfo{title}{{D}ynamic {A}sset {A}llocation under {I}nflation}.
\newblock \bibinfo{journal}{The Journal of Finance} \bibinfo{volume}{57},
  \bibinfo{pages}{1201--1238}.
\bibitem[{Brown and Warshawsky(2013)}]{Br13}
\bibinfo{author}{Brown, J.}, \bibinfo{author}{Warshawsky, M.},
  \bibinfo{year}{2013}.
\newblock \bibinfo{title}{{T}he {L}ife {C}are {A}nnuity: a {N}ew {E}mpirical
  {E}xamination of an {I}nsurance {I}nnovation that {A}ddresses {P}roblems in
  the {M}arkets for {L}ife {A}nnuities and {L}ong-term {C}are {I}nsurance}.
\newblock \bibinfo{journal}{Journal of Risk and Insurance}
  \bibinfo{volume}{80}, \bibinfo{pages}{677--704}.
\bibitem[{Cairns et~al.(2008)Cairns, Blake and Dowd}]{Ca08}
\bibinfo{author}{Cairns, A.}, \bibinfo{author}{Blake, D.},
  \bibinfo{author}{Dowd, K.}, \bibinfo{year}{2008}.
\newblock \bibinfo{title}{{M}odelling and {M}anagement of {M}ortality {R}isk:
  {A} {R}eview}.
\newblock \bibinfo{journal}{Scandinavian Actuarial Journal}
  \bibinfo{volume}{2008}, \bibinfo{pages}{79--113}.
\bibitem[{Clare(2022)}]{Cl22}
\bibinfo{author}{Clare, R.}, \bibinfo{year}{2022}.
\newblock \bibinfo{title}{{D}evelopments in {A}ccount {B}alances}.
\newblock \bibinfo{publisher}{ASFA}.
\bibitem[{Davidoff et~al.(2005)Davidoff, Brown and Diamond}]{Da05}
\bibinfo{author}{Davidoff, T.}, \bibinfo{author}{Brown, J.},
  \bibinfo{author}{Diamond, P.}, \bibinfo{year}{2005}.
\newblock \bibinfo{title}{{A}nnuities and {I}ndividual {W}elfare}.
\newblock \bibinfo{journal}{American Economic Review} \bibinfo{volume}{95},
  \bibinfo{pages}{1573--1590}.
\bibitem[{De~Felice(2023)}]{De23}
\bibinfo{author}{De~Felice, L.}, \bibinfo{year}{2023}.
\newblock \bibinfo{title}{Optimal {S}trategies for the {D}ecumulation of
  {R}etirement {S}avings under {V}arying {R}isk {A}versions}.
\newblock \bibinfo{type}{Honours {T}hesis}. University of Melbourne.
\bibitem[{De~Waegenaere et~al.(2010)De~Waegenaere, Melenberg and
  Stevens}]{Wa10}
\bibinfo{author}{De~Waegenaere, A.}, \bibinfo{author}{Melenberg, B.},
  \bibinfo{author}{Stevens, R.}, \bibinfo{year}{2010}.
\newblock \bibinfo{title}{{L}ongevity {R}isk}.
\newblock \bibinfo{journal}{De Economist} \bibinfo{volume}{158},
  \bibinfo{pages}{151--192}.
\bibitem[{Denuit et~al.(2011)Denuit, Haberman and Renshaw}]{De11}
\bibinfo{author}{Denuit, M.}, \bibinfo{author}{Haberman, S.},
  \bibinfo{author}{Renshaw, A.}, \bibinfo{year}{2011}.
\newblock \bibinfo{title}{{L}ongevity-indexed {L}ife {A}nnuitie}.
\newblock \bibinfo{journal}{North American Actuarial Journal}
  \bibinfo{volume}{15}, \bibinfo{pages}{97--111}.
\bibitem[{Dufresne(1990)}]{Du90}
\bibinfo{author}{Dufresne, D.}, \bibinfo{year}{1990}.
\newblock \bibinfo{title}{{T}he {D}istribution of a {P}erpetuity, with
  {A}pplications to {R}isk {T}heory and {P}ension {F}unding}.
\newblock \bibinfo{journal}{Scandinavian Actuarial Journal}
  \bibinfo{volume}{1990}, \bibinfo{pages}{39--79}.
\bibitem[{Dus et~al.(2005)Dus, Maurer and Mitchell}]{DuMaMi05}
\bibinfo{author}{Dus, I.}, \bibinfo{author}{Maurer, R.},
  \bibinfo{author}{Mitchell, O.}, \bibinfo{year}{2005}.
\newblock \bibinfo{title}{{B}etting on {D}eath and {C}apital {M}arkets in
  {R}etirement: a {S}hortfall {R}isk {A}nalysis of {L}ife {A}nnuities}.
\newblock \bibinfo{publisher}{National Bureau of Economic Research Cambridge,
  Mass., USA}.
\bibitem[{Ellis(2022)}]{El22}
\bibinfo{author}{Ellis, L.}, \bibinfo{year}{2022}.
\newblock \bibinfo{title}{The {N}eutral {R}ate: {T}he {P}ole-star {C}asts
  {F}aint {L}ight}.
\newblock \URLprefix \url{https://www.rba.gov.au/speeches/2022}.
\bibitem[{Evans and Sherris(2010)}]{EvSh10}
\bibinfo{author}{Evans, J.}, \bibinfo{author}{Sherris, M.},
  \bibinfo{year}{2010}.
\newblock \bibinfo{title}{{L}ongevity {R}isk {M}anagement and the {D}evelopment
  of a {L}ife {A}nnuity {M}arket in {A}ustralia}.
\newblock Australian School of Business Research Paper No. 2010ACTL01.
\bibitem[{Finlay and Wende(2018)}]{Fi18}
\bibinfo{author}{Finlay, R.}, \bibinfo{author}{Wende, S.},
  \bibinfo{year}{2018}.
\newblock \bibinfo{title}{{E}stimating {I}nflation {E}xpectations with a
  {L}imited {N}umber of {I}nflation-indexed {B}onds}.
\newblock \bibinfo{journal}{International Journal of Central Banking}
  \bibinfo{volume}{29}.
\bibitem[{Frank and Blanchett(2010)}]{FrBl10}
\bibinfo{author}{Frank, L.}, \bibinfo{author}{Blanchett, D.},
  \bibinfo{year}{2010}.
\newblock \bibinfo{title}{{T}he {D}ynamic {I}mplications of {S}equence {R}isk
  on a {D}istribution {P}ortfolio}.
\newblock \bibinfo{journal}{Journal of Financial Planning}
  \bibinfo{volume}{23}, \bibinfo{pages}{52--61}.
\bibitem[{Frank et~al.(2022)Frank, Mitchell and Blanchett}]{FrMiBl10}
\bibinfo{author}{Frank, L.}, \bibinfo{author}{Mitchell, J.},
  \bibinfo{author}{Blanchett, D.}, \bibinfo{year}{2022}.
\newblock \bibinfo{title}{{S}equence {R}isk: {M}anaging {R}etiree {E}xposure to
  {S}equence {R}isk {T}hrough {P}robability of {F}ailure {B}ased {D}ecision
  {R}ules}.
\newblock \bibinfo{journal}{Journal of Insurance and Financial Management}
  \bibinfo{volume}{6}.
\bibitem[{Fullmer(2008)}]{Fu08}
\bibinfo{author}{Fullmer, R.}, \bibinfo{year}{2008}.
\newblock \bibinfo{title}{{T}he {F}undamental {D}ifferences in {A}ccumulation
  and {D}ecumulation}.
\newblock \bibinfo{journal}{Journal of Investment Consulting}
  \bibinfo{volume}{9}, \bibinfo{pages}{36--40}.
\bibitem[{Hawley and Er(2022)}]{IFA22}
\bibinfo{author}{Hawley, E.}, \bibinfo{author}{Er, C.}, \bibinfo{year}{2022}.
\newblock \bibinfo{title}{Identifying Opportunities for Innovation in
  Post-retirement Products}.
\newblock \bibinfo{publisher}{Institute and Faculty of Actuaries}.
\bibitem[{Horneff et~al.(2008)Horneff, Maurer and Stamos}]{Ho08}
\bibinfo{author}{Horneff, W.}, \bibinfo{author}{Maurer, R.H.},
  \bibinfo{author}{Stamos, M.}, \bibinfo{year}{2008}.
\newblock \bibinfo{title}{{O}ptimal {G}radual {A}nnuitization: {Q}uantifying
  the {C}osts of {S}witching to {A}nnuities}.
\newblock \bibinfo{journal}{Journal of Risk and Insurance}
  \bibinfo{volume}{75}, \bibinfo{pages}{1019--1038}.
\bibitem[{Hurd and Rohwedder(2008)}]{Hu08}
\bibinfo{author}{Hurd, M.D.}, \bibinfo{author}{Rohwedder, S.},
  \bibinfo{year}{2008}.
\newblock \bibinfo{title}{The {R}etirement {C}onsumption {P}uzzle: {A}ctual
  {S}pending {C}hange in {P}anel {D}ata}.
\newblock \bibinfo{type}{Technical Report}. National Bureau of Economic
  Research.
\bibitem[{Khemka et~al.(2023)Khemka, Lim and Donnelly}]{DoKhLi23}
\bibinfo{author}{Khemka, G.}, \bibinfo{author}{Lim, W.},
  \bibinfo{author}{Donnelly, C.}, \bibinfo{year}{2023}.
\newblock \bibinfo{title}{{A}ligning {R}etirement {S}aving {G}oals with
  {I}nflation}.
\newblock \bibinfo{journal}{SSRN Electronic Journal} .
\bibitem[{Landriault et~al.(2018)Landriault, Li, Li and Young}]{La18}
\bibinfo{author}{Landriault, D.}, \bibinfo{author}{Li, B.},
  \bibinfo{author}{Li, D.}, \bibinfo{author}{Young, V.R.},
  \bibinfo{year}{2018}.
\newblock \bibinfo{title}{{E}quilibrium {S}trategies for the {M}ean-variance
  {I}nvestment {P}roblem over a {R}andom {H}orizon}.
\newblock \bibinfo{journal}{SIAM Journal on Financial Mathematics}
  \bibinfo{volume}{9}, \bibinfo{pages}{1046--1073}.
\bibitem[{Lee and Carter(1992)}]{LeCa92}
\bibinfo{author}{Lee, R.}, \bibinfo{author}{Carter, L.}, \bibinfo{year}{1992}.
\newblock \bibinfo{title}{{M}odeling and {F}orecasting {US} {M}ortality}.
\newblock \bibinfo{journal}{Journal of the American Statistical Association}
  \bibinfo{volume}{87}, \bibinfo{pages}{659--671}.
\bibitem[{Liu(2013)}]{Li13}
\bibinfo{author}{Liu, X.}, \bibinfo{year}{2013}.
\newblock \bibinfo{title}{{A}nnuity {U}ncertainty with {S}tochastic {M}ortality
  and {I}nterest {R}ates}.
\newblock \bibinfo{journal}{North American Actuarial Journal}
  \bibinfo{volume}{17}, \bibinfo{pages}{136--152}.
\bibitem[{Lockwood(2012)}]{Lo12}
\bibinfo{author}{Lockwood, L.}, \bibinfo{year}{2012}.
\newblock \bibinfo{title}{{B}equest {M}otives and the {A}nnuity {P}uzzle}.
\newblock \bibinfo{journal}{Review of Economic Dynamics} \bibinfo{volume}{15},
  \bibinfo{pages}{226--243}.
\bibitem[{Lockwood(2018)}]{Lo14}
\bibinfo{author}{Lockwood, L.}, \bibinfo{year}{2018}.
\newblock \bibinfo{title}{{I}ncidental {B}equests and the {C}hoice to
  {S}elf-insure {L}ate-life {R}isks}.
\newblock \bibinfo{journal}{American Economic Review} \bibinfo{volume}{108},
  \bibinfo{pages}{2513--2550}.
\bibitem[{Markowitz(1952)}]{Ma52}
\bibinfo{author}{Markowitz, H.}, \bibinfo{year}{1952}.
\newblock \bibinfo{title}{{P}ortfolio {S}election}.
\newblock \bibinfo{journal}{The Journal of Finance} \bibinfo{volume}{7},
  \bibinfo{pages}{77--91}.
\bibitem[{Merton(1969)}]{Me69}
\bibinfo{author}{Merton, R.}, \bibinfo{year}{1969}.
\newblock \bibinfo{title}{{L}ifetime {P}ortfolio {S}election under
  {U}ncertainty: the {C}ontinuous-time {C}ase}.
\newblock \bibinfo{journal}{The Review of Economics and Statistics}
  \bibinfo{volume}{51}, \bibinfo{pages}{247--257}.
\bibitem[{Milevsky(1997)}]{Mi97}
\bibinfo{author}{Milevsky, M.}, \bibinfo{year}{1997}.
\newblock \bibinfo{title}{{T}he {P}resent {V}alue of a {S}tochastic
  {P}erpetuity and the {G}amma {D}istribution}.
\newblock \bibinfo{journal}{Insurance: Mathematics and Economics}
  \bibinfo{volume}{20}, \bibinfo{pages}{243--250}.
\bibitem[{Milevsky and Posner(1998)}]{Mi98a}
\bibinfo{author}{Milevsky, M.}, \bibinfo{author}{Posner, S.},
  \bibinfo{year}{1998}.
\newblock \bibinfo{title}{Asian {O}ptions, the {S}um of {L}ognormals, and the
  {R}eciprocal {G}amma {D}istribution}.
\newblock \bibinfo{journal}{Journal of Financial and Quantitative Analysis}
  \bibinfo{volume}{33}, \bibinfo{pages}{409--422}.
\bibitem[{Milevsky et~al.(2005)Milevsky, Promislow and Young}]{MiYo05}
\bibinfo{author}{Milevsky, M.}, \bibinfo{author}{Promislow, D.},
  \bibinfo{author}{Young, V.}, \bibinfo{year}{2005}.
\newblock \bibinfo{title}{{F}inancial {V}aluation of {M}ortality {R}isk via the
  {I}nstantaneous {S}harpe {R}atio}.
\newblock \bibinfo{publisher}{IFID Center, Toronto, Canada}.
\bibitem[{Milevsky et~al.(2006)Milevsky, Promislow and Young}]{MiPr06}
\bibinfo{author}{Milevsky, M.}, \bibinfo{author}{Promislow, D.},
  \bibinfo{author}{Young, V.}, \bibinfo{year}{2006}.
\newblock \bibinfo{title}{{K}illing the {L}aw of {L}arge {N}umbers: {M}ortality
  {R}isk {P}remiums and the {S}harpe {R}atio}.
\newblock \bibinfo{journal}{Journal of Risk and Insurance}
  \bibinfo{volume}{73}, \bibinfo{pages}{673--686}.
\bibitem[{Milevsky and Robinson(2005)}]{MiRo05}
\bibinfo{author}{Milevsky, M.}, \bibinfo{author}{Robinson, C.},
  \bibinfo{year}{2005}.
\newblock \bibinfo{title}{{A} {S}ustainable {S}pending {R}ate without
  {S}imulation}.
\bibitem[{Milevsky and Young(2007)}]{Mi07}
\bibinfo{author}{Milevsky, M.}, \bibinfo{author}{Young, V.},
  \bibinfo{year}{2007}.
\newblock \bibinfo{title}{{A}nnuitization and {A}sset {A}llocation}.
\newblock \bibinfo{journal}{Journal of Economic Dynamics and Control}
  \bibinfo{volume}{31}, \bibinfo{pages}{3138--3177}.
\bibitem[{Milevsky and Huang(2011)}]{MiHu11}
\bibinfo{author}{Milevsky, M.A.}, \bibinfo{author}{Huang, H.},
  \bibinfo{year}{2011}.
\newblock \bibinfo{title}{{S}pending {R}etirement on {P}lanet {V}ulcan: {T}he
  {I}mpact of {L}ongevity {R}isk {A}version on {O}ptimal {W}ithdrawal {R}ates}.
\newblock \bibinfo{journal}{Financial Analysts Journal} \bibinfo{volume}{67},
  \bibinfo{pages}{45--58}.
\bibitem[{Mindlin(2009)}]{Di09}
\bibinfo{author}{Mindlin, D.}, \bibinfo{year}{2009}.
\newblock \bibinfo{title}{The Case for Stochastic Present Values}.
\newblock \bibinfo{publisher}{Society of Actuaries}.
\bibitem[{Mitchell and McCarthy(2002)}]{Mi02}
\bibinfo{author}{Mitchell, O.S.}, \bibinfo{author}{McCarthy, D.},
  \bibinfo{year}{2002}.
\newblock \bibinfo{title}{Annuities for an {A}geing {W}orld} .
\bibitem[{Modigliani(1986)}]{Mo86}
\bibinfo{author}{Modigliani, F.}, \bibinfo{year}{1986}.
\newblock \bibinfo{title}{{L}ife {C}ycle, {I}ndividual {T}hrift, and the
  {W}ealth of {N}ations}.
\newblock \bibinfo{journal}{The American Economic Review} \bibinfo{volume}{76},
  \bibinfo{pages}{297--313}.
\bibitem[{O~Meara et~al.(2015)O~Meara, Sharma and Bruhn}]{o15}
\bibinfo{author}{O~Meara, T.}, \bibinfo{author}{Sharma, A.},
  \bibinfo{author}{Bruhn, A.}, \bibinfo{year}{2015}.
\newblock \bibinfo{title}{{A}ustralia's {P}iece of the {P}uzzle -- {W}hy
  {D}on't {A}ustralians {B}uy {A}nnuities?}
\newblock \bibinfo{journal}{Australian Journal of Actuarial Practice}
  \bibinfo{volume}{3}, \bibinfo{pages}{47--57}.
\bibitem[{Olivieri(2001)}]{Ol01}
\bibinfo{author}{Olivieri, A.}, \bibinfo{year}{2001}.
\newblock \bibinfo{title}{{U}ncertainty in {M}ortality {P}rojections: an
  {A}ctuarial {P}erspective}.
\newblock \bibinfo{journal}{Insurance: Mathematics and Economics}
  \bibinfo{volume}{29}, \bibinfo{pages}{231--245}.
\bibitem[{Peijnenburg et~al.(2017)Peijnenburg, Nijman and Werker}]{Pe17}
\bibinfo{author}{Peijnenburg, K.}, \bibinfo{author}{Nijman, T.},
  \bibinfo{author}{Werker, B.}, \bibinfo{year}{2017}.
\newblock \bibinfo{title}{{H}ealth {C}ost {R}isk: a {P}otential {S}olution to
  the {A}nnuity {P}uzzle}.
\newblock \bibinfo{journal}{The Economic Journal} \bibinfo{volume}{127},
  \bibinfo{pages}{1598--1625}.
\bibitem[{Pfau(2015)}]{Pf15}
\bibinfo{author}{Pfau, W.}, \bibinfo{year}{2015}.
\newblock \bibinfo{title}{{O}ptimizing {R}etirement {I}ncome by {C}ombining
  {A}ctuarial {S}cience and {I}nvestments}.
\newblock \bibinfo{journal}{Retirement Management Journal} \bibinfo{volume}{5}.
\bibitem[{Piggott et~al.(2005)Piggott, Valdez and Detzel}]{Pi05}
\bibinfo{author}{Piggott, J.}, \bibinfo{author}{Valdez, E.},
  \bibinfo{author}{Detzel, B.}, \bibinfo{year}{2005}.
\newblock \bibinfo{title}{{T}he {S}imple {A}nalytics of a {P}ooled {A}nnuity
  {F}und}.
\newblock \bibinfo{journal}{Journal of Risk and Insurance}
  \bibinfo{volume}{72}, \bibinfo{pages}{497--520}.
\bibitem[{Pitacco(2007)}]{Pi07}
\bibinfo{author}{Pitacco, E.}, \bibinfo{year}{2007}.
\newblock \bibinfo{title}{Mortality and {L}ongevity: a {R}isk {M}anagement
  {P}erspective}.
\newblock \bibinfo{publisher}{IAA Life Colloquium}.
\bibitem[{Pitacco(2016)}]{Pi16}
\bibinfo{author}{Pitacco, E.}, \bibinfo{year}{2016}.
\newblock \bibinfo{title}{{G}uarantee {S}tructures in {L}ife {A}nnuities: a
  {C}omparative {A}nalysis}.
\newblock \bibinfo{journal}{The Geneva Papers on Risk and Insurance - Issues
  and Practice} \bibinfo{volume}{41}, \bibinfo{pages}{78--97}.
\bibitem[{Pitacco and Tabakova(2022)}]{PiTa22}
\bibinfo{author}{Pitacco, E.}, \bibinfo{author}{Tabakova, D.},
  \bibinfo{year}{2022}.
\newblock \bibinfo{title}{{S}pecial-rate {L}ife {A}nnuities: {A}nalysis of
  {P}ortfolio {R}isk {P}rofiles}.
\newblock \bibinfo{journal}{Risks} \bibinfo{volume}{10}.
\bibitem[{Powers(1995)}]{Po95}
\bibinfo{author}{Powers, M.}, \bibinfo{year}{1995}.
\newblock \bibinfo{title}{{A} {T}heory of {R}isk, {R}eturn and {S}olvency}.
\newblock \bibinfo{journal}{Insurance: Mathematics and Economics}
  \bibinfo{volume}{17}, \bibinfo{pages}{101--118}.
\bibitem[{Ramsay and Oguledo(2018)}]{RaOg18}
\bibinfo{author}{Ramsay, C.}, \bibinfo{author}{Oguledo, V.},
  \bibinfo{year}{2018}.
\newblock \bibinfo{title}{{T}he {A}nnuity {P}uzzle and an {O}utline of {I}ts
  {S}olution}.
\newblock \bibinfo{journal}{North American Actuarial Journal}
  \bibinfo{volume}{22}, \bibinfo{pages}{623--645}.
\bibitem[{RBA(2023)}]{rb22}
\bibinfo{author}{RBA}, \bibinfo{year}{2023}.
\newblock \bibinfo{title}{{I}nflation {T}arget}.
\newblock \bibinfo{howpublished}{\url{https://www.rba.gov.au/inflation/}}.
\bibitem[{Reichenstein and Dorsett(1995)}]{Re95}
\bibinfo{author}{Reichenstein, W.}, \bibinfo{author}{Dorsett, D.},
  \bibinfo{year}{1995}.
\newblock \bibinfo{title}{{T}ime {D}iversification {R}evisited}.
\newblock \bibinfo{publisher}{Research Foundation of the Institute of Chartered
  Financial Analysts}.
\bibitem[{Rice(2014)}]{Ri14}
\bibinfo{author}{Rice, M.}, \bibinfo{year}{2014}.
\newblock \bibinfo{title}{Investing for the {D}ifferent {P}hases of
  {R}etirement}.
\newblock \URLprefix
  \url{https://www.actuaries.digital/2014/09/15/investing-for-the-different-phases-of-retirement/}.
\bibitem[{Ritholz(2017)}]{Sh17}
\bibinfo{author}{Ritholz, B.}, \bibinfo{year}{2017}.
\newblock \bibinfo{title}{Tackling the `Nastiest, Hardest Problem in Finance'}.
\newblock \bibinfo{publisher}{Bloomberg}.
\bibitem[{Sheshinski(2008)}]{Sh08}
\bibinfo{author}{Sheshinski, E.}, \bibinfo{year}{2008}.
\newblock \bibinfo{title}{The Economic Theory of Annuities}.
\newblock \bibinfo{publisher}{Princeton University Press}.
\bibitem[{Thorburn(2018)}]{AGA}
\bibinfo{author}{Thorburn, G.}, \bibinfo{year}{2018}.
\newblock \bibinfo{title}{Retirement Income Risk Measure}.
\newblock \bibinfo{publisher}{Australian Government Actuary}.
\bibitem[{Thorburn(2019)}]{AGALT}
\bibinfo{author}{Thorburn, G.}, \bibinfo{year}{2019}.
\newblock \bibinfo{title}{{A}ustralian {L}ife {T}ables 2015-17}.
\newblock
  \bibinfo{howpublished}{\url{https://aga.gov.au/publications/life-tables/australian-life-tables-2015-17}}.
\bibitem[{Yaari(1965)}]{Ya65}
\bibinfo{author}{Yaari, M.}, \bibinfo{year}{1965}.
\newblock \bibinfo{title}{{U}ncertain {L}ifetime, {L}ife {I}nsurance, and the
  {T}heory of the {C}onsumer}.
\newblock \bibinfo{journal}{The Review of Economic Studies}
  \bibinfo{volume}{32}, \bibinfo{pages}{137--150}.
\bibitem[{Yin(2023)}]{Yi23}
\bibinfo{author}{Yin, Y.}, \bibinfo{year}{2023}.
\newblock \bibinfo{title}{The {N}ational {R}etirement {R}isk {I}ndex: {V}ersion
  2.0} .
\bibitem[{Zhang(2021)}]{Zh21}
\bibinfo{author}{Zhang, Z.}, \bibinfo{year}{2021}.
\newblock \bibinfo{title}{{S}tock {R}eturns and {I}nflation {R}edux: an
  {E}xplanation from {M}onetary {P}olicy in {A}dvanced and {E}merging
  {M}arkets}.
\newblock \bibinfo{publisher}{International Monetary Fund}.
\bibitem[{Zhou(2020)}]{Zh20}
\bibinfo{author}{Zhou, L.}, \bibinfo{year}{2020}.
\newblock \bibinfo{title}{A {S}tructured {I}nvestigation of {R}etirement
  {I}ncome {P}roducts}.
\newblock \bibinfo{publisher}{Australian Prudential Regulatory Authority}.
\bibitem[{Zhou et~al.(2022)Zhou, Sherris, Ziveyi and Xu}]{ZhShXu22}
\bibinfo{author}{Zhou, Y.}, \bibinfo{author}{Sherris, M.},
  \bibinfo{author}{Ziveyi, J.}, \bibinfo{author}{Xu, M.}, \bibinfo{year}{2022}.
\newblock \bibinfo{title}{{A}n {I}nnovative {D}esign of {F}lexible,
  {B}equest-enhanced {L}ife annuity with {N}atural {H}edging}.
\newblock \bibinfo{journal}{Scandinavian Actuarial Journal}
  \bibinfo{volume}{2022}, \bibinfo{pages}{488--509}.
\bibitem[{Zhou-Richter and Gr{\"u}ndl(2011)}]{Zh11}
\bibinfo{author}{Zhou-Richter, T.}, \bibinfo{author}{Gr{\"u}ndl, H.},
  \bibinfo{year}{2011}.
\newblock \bibinfo{title}{Life Care Annuities: Trick or Treat for Insurance
  Companies?}
\newblock Number \bibinfo{number}{04/11} in \bibinfo{series}{ICIR Working Paper
  Series}.

\end{thebibliography}

\appendix
\counterwithin{figure}{section}
\counterwithin{table}{section}

\section{Appendix: Natural Hedging Implementation} \label{A_nathed}
We introduced the concept of natural hedging in Section \ref{sec:nat_hedge} within the context of the Annuity Family. We now share some details about the implementation of natural hedging in \citet{De23}. Based on (\ref{eq:gsabe}), the set of admissible strategies will now be defined over $(w, \phi, \beta)$ and $(g(t))_{t\in \mathcal{S}}$.

\subsection{Effect of Natural Hedging on Loadings} We first generalise the individual cost function to permit natural hedging across the Annuity Family.
\begin{equation}\label{eq:icf}
C_{\angl{T_x}} = \int_0^{T_x} \phi I(t;\tilde{\mathbf{u}})g(t)\dd{t} + \phi\beta I(T_x, \tilde{\mathbf{u}})g(T_x).
\end{equation}
We only have $\theta > 0$ for the IIA and ULA  based on discussion in Section \ref{sec:KE}. For some trading strategy $\mathbf{z}$, we can define the discount factor $v_{\mathbf{z}} = \mathbb{E}[I(t; \mathbf{z})]$, and the actuarial present value  $a_{\angl{T}\mathbf{z}}$ of a continuous annuity over $[0, T]$. Denoting $v_{\mathbf{z}}= \mathrm{e}^{-\tilde{z}}$, the following identity holds for any $T\geq 0$:
\begin{equation}\label{eq:id}
1 = v_{\mathbf{z}}^{T} + \tilde{z}a_{\angl{T}\mathbf{z}}.
\end{equation}
For the IIA and ULA, we substitute for $g(t)$ in (\ref{eq:icf}) to write
\begin{equation}\label{eq:iden}
C_{\angl{T_x}} = \int_0^{T_x} \phi \mathbb{E}[I(t;\tilde{\mathbf{u}})] \dd{t} + \phi \beta \mathbb{E}[I(T_x; \tilde{\mathbf{u}})\vert T_x].
\end{equation}
Denoting $v_{\tilde{\mathbf{u}}}= \mathrm{e}^{-\tilde{u}}$, we have $\tilde{u} = r$ for the IIA and $\tilde{u} = \mu(w)$ for the ULA based on the restrictions on $\tilde{\mathbf{u}}$ imposed in Section \ref{sec:dec_v}. In general, we can draw on (\ref{eq:id}) to write (\ref{eq:iden}) as follows: 
\begin{equation}\label{eq:iden1}
C_{\angl{T_x}} = \phi(1 - \beta \tilde{u}) a_{\angl{T_x}\tilde{\mathbf{u}}}.
\end{equation}
If we are to consider a death benefit strictly for the purpose of natural hedging, then we can permit the retiree to inject an additional amount $A_{x;\tilde{\mathbf{u}}}\coloneqq\mathbb{E}[v_{\tilde{\mathbf{u}}}^{T_x}]$ at time $0$, maintaining the same actuarially fair price $P_0$ as under $\beta = 0$. As $\beta$ rises, we can average $C_{\angl{T_x}}$ over the pool to show in (\ref{eq:var_red}) that the variance of the average cost function decreases.
\begin{equation}\label{eq:var_red}
\mathrm{Var}(C_{x;\beta}^{(n)} ) = (1 - \beta \tilde{u})^2 \mathrm{Var}(C_{x;0}^{(n)} ).
\end{equation}
It follows from the loading assumption around (\ref{eq:sh0}) that $\theta_{\beta}\leq \theta_0$ and $\theta_{\beta}$ decreases in $\beta$. That is, 
\begin{equation}\label{eq:load_red}
\theta_{\beta} = (1 - \beta \tilde{u})\theta_0.
\end{equation}
Under the restriction that $\theta_{\beta} \geq 0$, we can also impose a bound such that $\beta\tilde{u}\leq 1$. \subsection{Effect of Natural Hedging on Theoretical Results}
Natural hedging will impact the components of the optimisation problem through both reducing loadings and changing the pattern of cash flows. By permitting the retiree to undertake natural hedging, (\ref{eq:gsa_eq}) generalises as follows:
\begin{equation}\label{eq:Xb}
1 -  (1 + \theta_{\beta})P_0 - \phi \beta A_{x;\tilde{\mathbf{u}}} = X_{\angl{T}}+ \int_0^{T} I(t; \mathbf{u})\big(c(t) - \phi \hat{g}(t)\big)\dd{t} - \phi \beta I(T_x, \tilde{\mathbf{u}}) \hat{g}(T_x)\mathbbm{1}_{\{ T_x \leq T\}}.\end{equation}
\subsubsection{Expected Investment Utility}
We generalise (\ref{eq:XI0}) to permit natural hedging by noting that the probability of death over interval $[t,t+\dd{t}]$ is $\lambda \mathrm{e}^{-\lambda t}$. We can therefore apply the $\mathcal{I}$ information set to (\ref{eq:Xb}) for $T = T_x$ as follows:
\begin{equation}\label{eq:Xb1}
1 -  (1 + \theta_{\beta})P_0 - \phi \beta A_{x;\tilde{\mathbf{u}}} = X_{x}^{\mathcal{I}}+ \int_0^{\infty} \mathrm{e}^{-\lambda t} I(t; \mathbf{u})\big(c(t) - \phi(1 + \beta \lambda) \hat{g}(t)\big)\dd{t}.\end{equation}

\subsubsection{Liquidity Shortfall Probability}
We generalise (\ref{eq:nu}) to permit natural hedging by noting that any lump sum received at $T_x$ will almost surely have no impact on the liquidity shortfall probability over $[0, T_x^*]$. We can therefore find the time of first shortfall as equal to the smallest \rev{non-negative} root $T = \tau$ of (\ref{eq:nu1}):
\begin{equation}\label{eq:nu1} \nu\, =\, 1 - (1 + \theta_{\beta}) P_{0} -  \phi \beta A_{x;\tilde{\mathbf{u}}}- \int_0^{T} \Big(c - \phi(1+\beta\lambda)\xi(t) \dfrac{_tp_x}{_tP_x^{(n)}}\Big)\mathrm{e}^{-\mu(w)t}\dd{t}.\end{equation}

\end{document}